\documentclass[format=acmsmall, review=false, screen=true]{acmart}

\usepackage{booktabs} % For formal tables
\usepackage{amsmath}
\usepackage{subcaption}
\usepackage{tablefootnote}
\usepackage{multirow}
\usepackage{graphicx}
\usepackage[flushleft]{threeparttable}
\usepackage{url}
\usepackage{xspace}
\usepackage{refcount}

\usepackage[ruled]{algorithm2e} % For algorithms

\SetAlFnt{\small}
\SetAlCapFnt{\small}
\SetAlCapNameFnt{\small}
\SetAlCapHSkip{0pt}
\IncMargin{-\parindent}

\setcopyright{acmlicensed}
\acmDOI{0000001.0000001}

\received{June 2017}
\received[revised]{September 2018}
\received[accepted]{December 2018}

\usepackage{color}
\definecolor{darkgreen}{HTML}{008800}

\newcommand{\aasustainable}{\textit{Sustainable}\xspace}
\newcommand{\aatransitioning}{\textit{Transitioning}\xspace}
\newcommand{\aaemerging}{\textit{Emerging}\xspace}

\newcommand{\aanonrecurring}{\textit{Non-Recurring Activity Archetype}\xspace}
\newcommand{\aasporadic}{\textit{Sporadic Activity Archetype}\xspace}
\newcommand{\aafrequent}{\textit{Frequent Activity Archetype}\xspace}
\newcommand{\aapermanent}{\textit{Permanent Activity Archetype}\xspace}
\newcommand{\aat}{\textit{Activity Archetype}\xspace}
\newcommand{\aats}{\textit{Activity Archetypes}\xspace}
\newcommand{\aaac}{\textit{activity composition}\xspace}
\newcommand{\aaacs}{\textit{activity compositions}\xspace}

\newcounter{tsacounter}
\DeclareRobustCommand{\tsa}[1]{\textbf{/* #1 (tsa) */}\stepcounter{tsacounter}\typeout{LaTeX Warning: tsa comment \thetsacounter: #1 (line \the\inputlineno)}}

\newcounter{swacounter}
\DeclareRobustCommand{\swa}[1]{\textbf{/* #1 (swa) */}\stepcounter{swacounter}\typeout{LaTeX Warning: swa comment \theswacounter: #1 (line \the\inputlineno)}}

\newcounter{dhecounter}
\DeclareRobustCommand{\dhe}[1]{\textbf{/* #1 (dhe) */}\stepcounter{dhecounter}\typeout{LaTeX Warning: dhe comment \thedhecounter: #1 (line \the\inputlineno)}}

\newcounter{rkocounter}
\DeclareRobustCommand{\rko}[1]{\textbf{/* #1 (rko) */}\stepcounter{rkocounter}\typeout{LaTeX Warning: rko comment \therkocounter: #1 (line \the\inputlineno)}}

\newcounter{mstcounter}
\DeclareRobustCommand{\mst}[1]{\textbf{/* #1 (mst) */}\stepcounter{mstcounter}\typeout{LaTeX Warning: mst comment \themstcounter: #1 (line \the\inputlineno)}}

\begin{document}
\title[Activity Archetypes in Q\&A Websites]{Activity Archetypes in Question-and-Answer (Q\&A) Websites - A Study of 50 Stack Exchange Instances}  
\author{Tiago Santos}
%\orcid{1234-5678-9012-3456}
\affiliation{%
  \institution{Graz University of Technology}
  \city{Graz}
  \country{Austria}}
\author{Simon Walk}
\affiliation{%
  \institution{Detego GmbH}
  \city{Graz}
  \country{Austria}}
\author{Roman Kern}
\affiliation{%
  \institution{Graz University of Technology \& Know-Center, Graz, Austria}
  \city{Graz}
  \country{Austria}}
\author{Markus Strohmaier}
\affiliation{%
  \institution{RWTH Aachen University \& GESIS}}
\author{Denis Helic}
\affiliation{%
  \institution{Graz University of Technology}}

\begin{abstract}
    Millions of users on the Internet discuss a variety of topics on Question-and-Answer (Q\&A) instances. However, not all instances and topics receive the same amount of attention, as some thrive and achieve self-sustaining levels of activity, while others fail to attract users and either never grow beyond being a small niche community or become inactive.
    Hence, it is imperative to not only better understand but also to distill deciding factors and rules that define and govern sustainable Q\&A instances.
    We aim to empower community managers with quantitative methods for them to better understand, control and foster their communities, and thus contribute to making the Web a more efficient place to exchange information.
    To that end, we extract, model and cluster user activity-based time series from $50$ randomly selected Q\&A instances from the Stack Exchange network to characterize user behavior.
    We find four distinct types of user activity temporal patterns, which vary primarily according to the users' activity frequency.
    Finally, by breaking down total activity in our $50$ Q\&A instances by the previously identified user activity profiles, we classify those $50$ Q\&A instances into three different activity profiles.
    Our parsimonious categorization of Q\&A instances aligns with the stage of development and maturity of the underlying communities, and can potentially help operators of such instances: We not only quantitatively assess progress of Q\&A instances, but we also derive practical implications for optimizing Q\&A community building efforts, as we e.g. recommend which user types to focus on at different developmental stages of a Q\&A community.
\end{abstract}

\begin{CCSXML}
 <ccs2012>
 <concept>
 <concept_id>10002950.10003648.10003688.10003697</concept_id>
 <concept_desc>Mathematics of computing~Cluster analysis</concept_desc>
 <concept_significance>500</concept_significance>
 </concept>
 <concept>
 <concept_id>10002950.10003648.10003688.10003693</concept_id>
 <concept_desc>Mathematics of computing~Time series analysis</concept_desc>
 <concept_significance>300</concept_significance>
 </concept>
 <concept>
 <concept_id>10002951.10003260.10003282.10003296.10003297</concept_id>
 <concept_desc>Information systems~Answer ranking</concept_desc>
 <concept_significance>500</concept_significance>
 </concept>
 <concept>
 <concept_id>10003120.10003121.10003124.10010868</concept_id>
 <concept_desc>Human-centered computing~Web-based interaction</concept_desc>
 <concept_significance>300</concept_significance>
 </concept>
 <concept>
 <concept_id>10003120.10003130.10003131.10003570</concept_id>
 <concept_desc>Human-centered computing~Computer supported cooperative work</concept_desc>
 <concept_significance>300</concept_significance>
 </concept>
 </ccs2012>
\end{CCSXML}

\ccsdesc[500]{Mathematics of computing~Cluster analysis}
\ccsdesc[300]{Mathematics of computing~Time series analysis}
\ccsdesc[500]{Information systems~Answer ranking}
\ccsdesc[300]{Human-centered computing~Web-based interaction}
\ccsdesc[300]{Human-centered computing~Computer supported cooperative work}


\keywords{Question-and-Answer (Q\&A) websites; User types in Q\&A websites; Temporal activity patterns in Q\&A websites; Sustainability of Q\&A websites}

\maketitle

%\renewcommand{\shortauthors}{G. Zhou et al.}

\section{Introduction}
Question-and-answer (Q\&A) websites (e.g., Stack Exchange\footnote{\url{http://www.stackexchange.com}} or Quora\footnote{\url{http://www.quora.com}}) are publicly accessible platforms, which are used by millions of users to discuss a variety of topics and problems. For example, the StackOverflow\footnote{\url{http://stackoverflow.com/}} instance of the Stack Exchange website deals with topics related to programming and hosts a flourishing community of more than $6$ million users. Another prominent example is the Math Stack Exchange\footnote{\url{http://math.stackexchange.com/}} instance, where a thriving community of mathematical professionals and other users with shared interests pose and solve mathematical questions.

\noindent \textbf{Problem.}
However, not all Q\&A instances exhibit the same kind of vibrant, self-sustaining community activity.
In fact, the majority of Q\&A instances fail to attract and engage enough users to reach self-sustainability in terms of activity.
Typically, instance operators provide incentives for users in the form of badges or reputation scores.
Although several studies analyzed the effects of such endeavors \cite{mamykina2011design, sinha2013exploring, anderson2013steering}, our research community still lacks the tools to understand, measure, model and predict key factors that influence and drive Q\&A communities to sustainable levels of activity.
However, without a proper understanding of users, the structures inherent in the communities, as well as the driving mechanisms behind successful Q\&A instances, we can not hope to remedy the problems of less successful sites.

\noindent \textbf{Approach.}
In this paper we set out (i) to characterize user activity profiles, (ii) to reveal the compositions of those profiles in various Q\&A communities, and (iii) to analyze similarities and differences between highly and less successful Q\&A instances.

Although current research on users of online Q\&A communities partially uncovers different user roles in these communities \cite{mamykina2011design, danescu2013no, furtado2013contributor, yang2014sparrows}, we identify a research gap on (i) the composition of activity profiles for communities at different stages of maturity and (ii) specific compositions that ultimately make thriving communities successful.

Specifically, we characterize temporal activity patterns of users of Q\&A instances, analyze and compare the activity composition and development of whole instances, and provide actionable information for instance operators to assess maturity, improve activity and manage their instances more efficiently.
To that end, we randomly pick a total of $50$ Stack Exchange instances, from which we derive time series and features which describe commonly occurring temporal activity patterns. We represent user activity as their total count of posts and replies.
Subsequently, we apply K-Means on the extracted features to group users with similar activity profiles and find optimal numbers of clusters by calculating and comparing silhouette coefficients for different values of K. 
Additionally, we analyze the composition of activity across the obtained clusters for all Q\&A instances. 

\noindent \textbf{Contributions.}
The main contributions of our work are as follows:
First, we find that activity-based time series can be described by the following two quantities: (i) the characteristics of its peaks and (ii) the uniqueness of its non-zero activity values.

Second, we identify typical user activity profiles to describe \textit{Activity Archetypes}, which represent distinct user engagement levels across all analyzed Stack Exchange instances.
This result helps not only to better understand the different user profiles that operators of Q\&A instances need to cater to, but also which profiles to include when modeling activity for these instances. 

Third, we analyze, compare and categorize the \aat composition of various Stack Exchange instances, which allows us to assess the level of maturity in a Stack Exchange instance's development towards activity-based self-sustainability.
To give an example, we find that thriving instances feature substantial amounts of infrequently active users posing questions.
If this group of users is underrepresented, then this affects the instance's overall activity and development.

We believe that our analyses represent an important step towards a better understanding of the factors that define and foster success in Q\&A instances.
With our analyses, we enable Q\&A instance operators not only to gauge, quantify and model the status of their communities in comparison to other communities, but, with our discussion of practical implications, also to pinpoint what user groups to focus activity improvement measures on, on the path towards a thriving, self-sustaining community.

\section{Related Work}
\noindent \textbf{Dynamical systems for modeling activity.}
Dynamical systems are systems of parametrized equations describing the evolution of numerical quantities over time.
They provide a mathematical formalization for activity dynamics models.

Perra et al. \cite{perra2012activity} model activity in collaboration networks such as publications and references in the Physical Review Letters journal.
The authors measure an empirical probability distribution over interactions of agents in a network, model the formation of dynamic networks based on this activity distribution and study resulting dynamical processes.
This work influenced other authors modeling activity dynamics as explicit dynamic processes on networks, such as Laurent et al. \cite{laurent2015calls}.
Those authors propose an activity-driven model for time varying networks to analyze mobile call records from an European telecom.
Building on the work by Perra et al. and Laurent et al., W{\"o}lbitsch et al.~\cite{wolbitsch2017modeling} extended an activity-driven network model with a peer-influence mechanism to study peer-influence and its effects on the network in a controlled setting.

Other approaches to model activity in Q\&A instances and networks with dynamical systems focus on a few key variables that drive overall activity dynamics.
Ribeiro \cite{ribeiro2014modeling} models activity in membership-based community websites as time series counting the number of active users in such websites.
The model considers two main factors, namely active users spontaneously becoming inactive and active users spurring inactive ones to become active.
These factors are sufficient to distinguish self-sustaining from non-self-sustaining online communities and to forecast their daily active user numbers.
Walk et al. \cite{walk2016activity} proposed a dynamical system description for online Q\&A instances such as Stack Exchange instances or Semantic MediaWikis\footnote{\url{https://www.semantic-mediawiki.org/wiki/Semantic_MediaWiki}}.
Their dynamical system equations allow for (i) forecasting activity levels in those online Q\&A communities, and for (ii) assessing if an online community reached self-sustaining levels of activity.
In an extension of Walk et al.'s~\cite{walk2016activity} models, Koncar et al.~\cite{koncar2017exploring} recently studied the implications of trolling behavior on various Stack Exchange and Reddit communities.

Similarly to our previous work~\cite{santos2017nonlinear} on nonlinear characterization of Q\&A instances, we contribute a data-driven approach to this body of work, which uses mathematical formalization to describe activity in online Q\&A instances.
In an extension of our previous work, however, we go beyond our analysis of time series of Q\&A activity totals by focusing on more granular activity-based time series.
Specifically, our objects of study in this work are time series describing user activity in Q\&A forums.
We thus empirically identify user behavior patterns as key driving forces of activity and thereby pave the way for new models, which take into account users' roles in shaping total Q\&A activity as it changes over time.

\noindent \textbf{Characterization of activity in Q\&A instances.}
Literature dealing with dynamics of Q\&A instances such as Stack Exchange focuses on many different aspects of these types of online communities.
Anderson et al. \cite{anderson2012discovering} quantify and uncover temporal characteristics of questions which bring (long-term) value to the community.
Burel and He \cite{burel2013question} measure the maturity of the ServerFault Stack Exchange instance by its ability to cope with complex questions.
In contrast to that study on complex questions, Correa and Sureka~\cite{correa2013fit, correa2014chaff} measure, via characterization studies and prediction experiments, the properties and impact of closed and deleted questions on Q\&A quality maintenance.
Srba et al. \cite{srba2015utilizing} aim to encourage activity on new questions in Stack Exchange instances with improvements on linking users to unanswered questions by analyzing a larger pool of data sources other than the Stack Exchange data itself (e.g. Twitter).
Our work also derives policy suggestions for Q\&A community managers, but from a user-based analysis, rather than based on questions and their properties.
This enables our focus on macro-level aspects of Q\&A community growth and management, complementing these more granular studies on the impact and value of questions.

Other authors, however, have, similarly to us, focused on user types and engagement as their fundamental object of study.
Danescu-Niculescu-Mizil et al. \cite{danescu2013no} characterize user participation in online communities by the evolution of their language, allowing the authors to predict when users depart their communities.
In another study of user types in an online Q\&A community, Gazan~\cite{gazan2006specialists, gazan2007seekers} highlights different types of questioners and answerers, namely seekers and sloths and respectively specialists and synthesists.
Zhang et al.~\cite{zhang2007expertise} and more recently Yang et al. \cite{yang2014sparrows} tackle the problem of expert user identification and characterization in, respectively, a help-seeking forum for Java programming and StackOverflow.
Put into a broader context, our work also relates to feature-based characterizations of user behavior online in general, such as Lehmann et al.'s work~\cite{lehmann2012models} on typifying online forums by their users' activity and Chan et al.'s work~\cite{chan2010decomposing} on user types and temporal aspects of user engagement in 80 websites.

Early work by Adamic et al.~\cite{adamic2008knowledge} and Nam et al.~\cite{nam2009questions} on understanding knowledge sharing behavior in the Yahoo Answers and Naver Q\&A communities explained user behavior as a product of users' interests and motivation.
More recently, Mamykina et al. \cite{mamykina2011design}'s analysis of the StackOverflow design combines a statistical investigation of StackOverflow usage patterns with interviews with StackOverflow's designers.
The goal of their procedure is to understand which user behavior leads to the site's success.
In particular, the authors find a set of three different types of user activity behavior (plus a lurker, non-active type), which base on their activity frequency.

Sinha et al. \cite{sinha2013exploring} study participation and participation incentives in Stack Exchange communities. 
In their work, the authors underline the relevance of a core of highly active users and of participation incentives for less active users in Stack Exchange communities.
Our work shares most commonalities with Furtado et al. \cite{furtado2013contributor}'s.
In that study, the authors extract metrics measuring quality and quantity of activity in Stack Exchange instances. 
With those metrics, they describe a set of ten different user profiles obtained with K-Means clustering on those extracted metrics.
The authors then study the composition and activity dynamics of users in five Stack Exchange instances broken down by the user profiles they found.
They show that, although users change profiles over time, the overall composition of user profiles of those five instances mostly does not.

Our comprehensive analysis of 50 Stack Exchange instances yields comparable, but, as we discuss later, slightly but crucially different user profile characterizations than those by Mamykina et al. \cite{mamykina2011design}, Sinha et al. \cite{sinha2013exploring} and Furtado et al. \cite{furtado2013contributor}.
That work provides the basis for our paper to expand on as follows: A temporal analysis of our user characterization enables us to uncover previously overseen patterns regarding the development and maturity of Stack Exchange instances of varying sizes, ages and activity profiles.
In particular, our results, which highlight an instance's evolving activity composition over time, do not contradict the findings by Furtado et al.~\cite{furtado2013contributor}.
We rather extend the results by Furtado et al.~\cite{furtado2013contributor}, as they analyzed only five similarly sized Stack Exchange instances, one of which (\textit{programmers}\footnote{As of February 2017, the \textit{programmers} Stack Exchange instance is termed \textit{softwareengineering} and \url{programmers.stackexchange.com} redirects to~\url{softwareengineering.stackexchange.com}.}) we find to be of one of multiple types we identify.
Our Q\&A user and instance characterization thus generalizes their work, as we uncover also a relation between not just one but several user compositions of Stack Exchange instances and their evolving activity growth.

We find that the works by Iriberri and Leroy~\cite{iriberri2009life} and by Young~\cite{young2013community} qualitatively corroborate our findings.
Those authors identify four main life-cycle phases of online communities, namely inception, establishment or growth, maturity and death or self-sustainability or mitosis, which are comparable to the Stack Exchange instance characterization we derive.
In particular, Young~\cite{young2013community} also derives a set of recommendations for online health community managers to adapt to their communities' different life-cycle stages. Similarly to Young, we also propose measures for boosting activity in Stack Exchange instances at different maturity stages.
In the context of the work by these authors, our work complements theirs with quantitative empirical results and with the application domain of online Q\&A communities.

We refer the interested reader to the survey by Srba and Bielikova~\cite{srba2016comprehensive} on previous work on community questions and answers websites for more literature on these topics.

\noindent \textbf{Time series clustering.}
In the task of time series clustering, one aims to group time series with similar shapes or properties, to ultimately categorize time series, find representative patterns and uncover hidden structures in time series.

A number of authors \cite{faloutsos1994fast, fu2001pattern, vlachos2003wavelet, liao2004clustering, hautamaki2008time, yang2011patterns} have applied time series clustering techniques to domains such as finance, online content spread, sensor data or even warfare analysis.
These authors share a common time series clustering approach, which begins with the choice of time series representation to feed to different clustering algorithms.
Authors such as Hautamaki et al. \cite{hautamaki2008time} consider time series without any transformation, while others extract features \cite{faloutsos1994fast,fu2001pattern} or apply transformations to the time series, such as Discrete Wavelet Transforms \cite{vlachos2003wavelet,yang2011patterns} and Symbolic Aggregate ApproXimation \cite{liao2004clustering}.
The time series clustering approach continues with the selection of a distance metric, which very often is the Euclidean \cite{faloutsos1994fast, vlachos2003wavelet, liao2004clustering} or the Dynamic Time Warping distance \cite{hautamaki2008time}.
Finally, authors settle on a time series clustering algorithm, with popular choices being K-Means and variations thereof \cite{vlachos2003wavelet, liao2004clustering, hautamaki2008time, yang2011patterns}, self-organizing maps \cite{fu2001pattern} and hierarchical clustering \cite{hautamaki2008time}. %minor_revisions
We select time series features and apply Euclidean K-Means on them, to cope with the challenge which discrete valued time series data presents and which is, according to Aghabozorgi et al.~\cite{aghabozorgi2015time}, rarely dealt with in time series clustering literature.
We encourage readers interested in more time series clustering methods and applications to acquaint themselves with the review by Aghabozorgi et al..

\section{Materials and Methods}
\subsection{Dataset Characterization}
\label{subsec:datasets}

We analyze questions and answers from $50$ Stack Exchange Q\&A instances on many diverse topics, such as \textit{tex}\footnote{\label{fnurl}All instances have a corresponding *.stackexchange.com website, where * denotes the instance's name.}, \textit{english}\footnotemark[\getrefnumber{fnurl}], \textit{gardening}\footnotemark[\getrefnumber{fnurl}] or \textit{buddhism}\footnotemark[\getrefnumber{fnurl}]. 
The observation periods for all instances vary between $4$ to $80$ months, depending on the inception date of each instance. The final observation month is February 2017.

\begin{figure*}[!t]
    \begin{subfigure}{.32\textwidth}
        \centering
        \includegraphics[width=5cm,height=5cm,keepaspectratio]{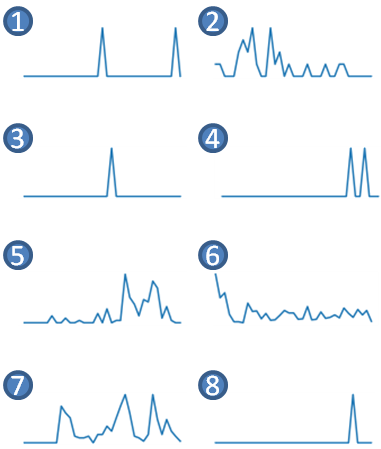}
        \caption{Extract time series}
        \label{fig:extract_timeseries}
    \end{subfigure}
    \begin{subfigure}{.32\textwidth}
        \centering
        \vspace*{11.3mm}
        \includegraphics[width=0.95\linewidth]{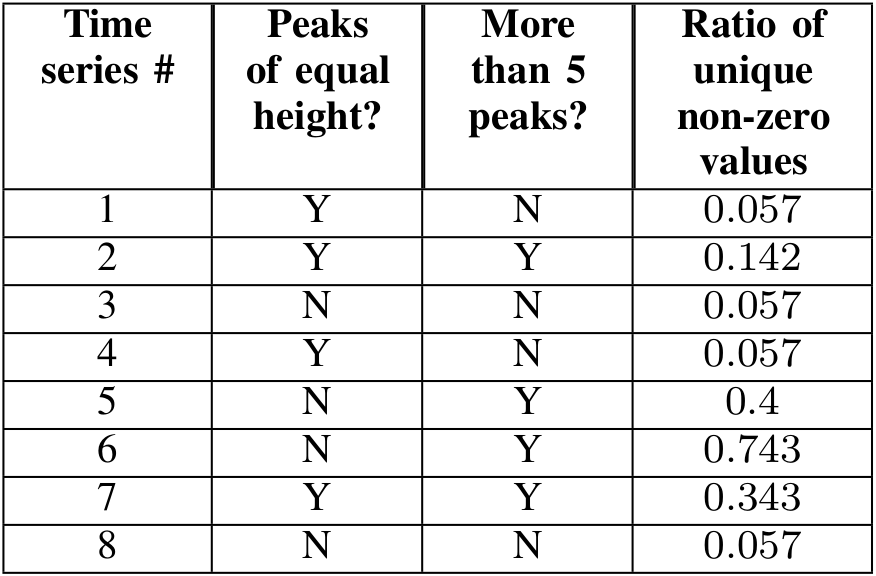}
        \vspace*{11.3mm}
        \caption{Compute features}
        \label{fig:compute_features}
    \end{subfigure}
    \begin{subfigure}{.33\textwidth}
        \centering
        %\vspace*{0.01mm}
        \includegraphics[width=1.03\linewidth]{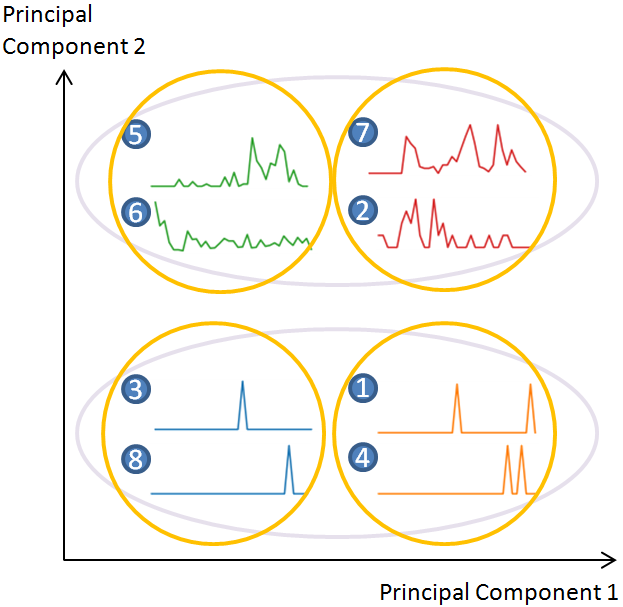}
        \vspace*{0.01mm}
        \caption{Search for optimal clustering}
        \label{fig:silhouette_maximization}
    \end{subfigure}
    \caption{\label{fig:timeSeriesClusteringProcess}
    \textbf{Identifying user archetypes as a time series clustering problem.} 
    We start by extracting questions-based (and, separately, answers-based) activity time series as the monthly sum of posted questions (respectively answers) of each user in a Stack Exchange instance (cf. Fig.~\ref{fig:extract_timeseries}).
    We then extract three features from these time series: two Boolean features, describing if an activity time series has peaks of equal maximal height and if it has more than five peaks, and the ratio of unique non-zero values to time series length, a continuous feature varying between zero and one (cf. Fig.~\ref{fig:compute_features}).
    Finally, we cluster the extracted features with K-Means for $K = 2, \ldots, 10$ and save $K^*$, the value of K which maximizes the average silhouette coefficient.
    Graphical inspection of the clusters via PCA projection to two-dimensional space (cf. Fig.~\ref{fig:silhouette_maximization}) yields well-separated and cohesive clusters for $K=2,3,4$.
    However, in this example, for $K=2,3,4$, we get average silhouette coefficient values of $0.423$, $0.563$ and $0.871$, respectively.
    Hence, $K^*$ equals four.
    In Stack Exchange instances, we observe varying $K^*$, which hints at different activity compositions in terms of user archetypes.
}
\end{figure*}

As different instances originate at different points in time, the communities in each of those instances naturally exhibit different levels of activity and maturity.
For example, \textit{english} started in June 2009 and attracted a total of $37,125$ users until February 2017. 
In contrast, \textit{earthscience}\footnotemark[\getrefnumber{fnurl}] managed to attract only $578$ users between April 2014 and February 2015.
To foster the development of young instances, such as \textit{earthscience}, the Stack Exchange community submits, incubates and evaluates proposals for new Q\&A instances at a dedicated website called Area 51\footnote{\url{http://area51.stackexchange.com/}}.
If an Area 51 Q\&A instance reaches a significant level of activity, the Area 51 community deems it ready for a live test. Then, its live deployment ensues and the Area 51 community monitors its progress until it reaches a sustainable level of activity. 

In this paper, we analyze a total of $50$ Stack Exchange instances consisting of $25$ randomly chosen Area 51 datasets and another $25$ randomly chosen non-Area 51 datasets (see Table~\ref{table:datasets}).

\begin{table}[!b]
    \centering
    \caption{\label{table:datasets}
    \textbf{Dataset characteristics.} We present value ranges for the number of users, activity (i.e. aggregated questions, answers and comments) and observation periods (in months) of all datasets (i.e. Stack Exchange instances) per dataset group. Instances listed on Area 51 are typically smaller and younger than those outside Area 51.
    }
    \begin{threeparttable}
    \begin{tabular}{l|c|c|c|c}
        \toprule
        \textbf{Dataset group}&\textbf{Size}&\textbf{Users}&\textbf{Activity}&\textbf{Months}\\
        \midrule
        Area 51&25&$[473, 6309]$&$[5023, 47421]$&$[4, 47]$\\
        \midrule
        Non-Area 51&25&$[1953, 37125]$&$[8137, 624166]$&$[11, 80]$\\
        \bottomrule
    \end{tabular}
    \end{threeparttable}
\end{table}

\subsection{Feature Engineering}
\label{sec:featureEng}

\noindent \textbf{Modeling user activity as time series.}
We model user activity in Stack Exchange online Q\&A instances as two activity-based time series per user. The first one comprises question counts, and the second one reply and comment counts for a given user per month. We stipulate that a user has zero activity if the user did not post a single question (or answer) in any given month of an instance's existence.
In all of the following, we treat questions-based activity time series separately from answers-based ones.

\noindent \textbf{Comparing users' activity-based time series directly.}
We aim to group users with similar activity profiles by clustering similar activity-based time series.

We first tried to base our clustering approach on a direct measure of similarity between users' activity-based time series with the Euclidean distance.
However, using the Euclidean distance fails to discern users with different activity profiles, as it does not account for the misalignment of activity bursts and other activity-affecting events. For example, notice the misalignment in the time axis of the activity peak in time series three and eight of Figure~\ref{fig:extract_timeseries}.
As a counter measure to compare misaligned time series, we employed Dynamic Time Warping (DTW).
DTW aligns time series over the time axis before computing their similarity with some measure such as Euclidean distance.
However, DTW lead to no improvements in activity-based time series clustering.
As Yang and Leskovec~\cite{yang2011patterns} point out, time series of comparable shape but overall varying volume would be considered similar by DTW, hence making the assignment of different time series into meaningful clusters harder.
We discarded other well-established time series similarity and clustering approaches, including SAX-based~\cite{keogh2004towards} and matrix profile-based~\cite{yeh2016matrix} approaches.
These and other times series clustering algorithms we reviewed do not specifically address the clustering of sparse count time series problem we face, and thus do not extract meaningful clusters.
Furthermore, we were also careful not to apply clustering to segmented time series (e.g. around user activity peaks) with distance-based metrics, as this may be problematic in practice~\cite{keogh2005clustering}.
The need for caution arises from clusters of time series subsequences being essentially random, i.e. independent of input time series subsequence types, if care is not taken to extract time series motifs rather than (often trivial) time series sliding windows.

\noindent \textbf{Extracting features describing temporal activity patterns.}
Hence, we devised a different approach: We extract time series features summarizing key aspects of temporal user activity patterns occurring in the $50$ Stack Exchange instances.

\textit{Feature selection.} 
To select time series features, we started with a list of more than $400$ different features~\cite{christ2016distributed}, which comprise descriptive statistics (e.g., mean, auto-correlation or kurtosis), time series models (e.g., auto-regressive coefficients), and other time series transformations (e.g., Fourier).
Starting with those features, we manually searched for a smaller and simpler set of features, which nicely capture the activity distributions as well as a given user activity behavior. Thereby we focused our search on the user behavior that we observed in our data or that was identified in the previous studies.
    
In general, we observe large numbers of users with sporadic peaks in activity as well as fewer users who are more active and contribute fluctuating amounts of questions and answers over longer periods of time (cf. examples in Figure~\ref{fig:extract_timeseries}). Thus, we observe so-called bursty patterns in user activity-based time series. Such bursty patterns have been frequently observed on the Web~\cite{mamykina2011design, vazquez2006modeling}.

Although the majority of our activity-based time series exhibited such patterns consistently, they varied in their temporal location.
In other words, rarely-active users with activity bursts in the beginning of the lifetime of a Stack Exchange instance behave similarly to rarely active users with activity bursts in the tail end of the same Stack Exchange instance.
This motivated our first filtering criterion for reducing the original set of more than $400$ time series features: We excluded locality based features, such as the locations of a time series' minimal values, since these types of features differentiate user behavior which we intuitively believe belongs grouped together.

Then, we excluded features unfit for modeling sparse count time series (i.e. our activity-based time series), such as continuous wavelet transformations, auto-regressive models and other descriptive statistics inadequate in a high sparseness context.
Instead, we focused on descriptive statistics which capture activity bursts, such as peak-related features (such as peak height or the number of peaks) or unique value counts.
These two filtering steps left us with a set of $15$ features.

We then empirically evaluated this set of 15 features with respect to parsimony, their descriptiveness of sparse and bursty user activity along the dimensions frequency and amplitude, and their contribution in a clustering experiment.
After this final feature selection step, we chose three activity-based time series features, which capture exactly these kinds of behavior: the ratio of unique non-zero values to time series length, a Boolean feature describing if the activity time series has more than five peaks and another Boolean feature measuring if the time series has maximum peaks of equal height.

\textit{Feature computation.} 
We compute activity peaks as values which are larger than other values in their direct neighborhood of the previous and next observations of the activity time series (cf. Figure~\ref{fig:compute_features}). 
Note that, with this definition of a peak, we do not impose a minimum peak height, both in absolute terms, as well as relative to the peak's neighboring values.
For example, assume a user posts a question once in January, responds twice to some other question in February and then asks one other question in March.
The activity-based time series corresponding to this user would thus feature one peak in February.
For the binary feature related to the number of peaks, we settled on the threshold $5$ (i.e. the feature measures if an activity-based time series has (or not) more than $5$ peaks). The reason for the threshold $5$ is that this threshold corresponds to the average 90th quantile of the number of peaks per activity-based time series. Thus, this feature separates a minority of users with high volumes of contributions (as measured by peaks in activity) from the majority of other users.

The other binary feature, checking if a user's activity-based time series has a duplicate maximum peak, captures regularity in behavior pattern, both of sparsely active users posting just two questions (or answers) in separate occasions or of regularly and frequently active users with consistently regular activity patterns.
Hence, this binary feature combines with the other one to separate users along the dimensions of activity volume over time and activity regularity.

Finally, the third feature, ratio of unique non-zero values to time series length, allows for finer shades of distinction between highly and regularly active from less and irregularly active users, as this continuous feature encompasses both sides of this spectrum.
On the one hand, frequently highly (sporadically less) active users will tend to have a low (high) such ratio, but variations of these two extremes are possible (e.g. regular and highly active users) and interesting for later analysis.
We use a ratio (and not the absolute count of unique non-zero values) to ensure all our features are defined over the interval $[0, 1]$, which allows for better comparison in Euclidean distance-based clustering methods.

\textit{Alternative feature extraction methods for clustering.} 
We note here that other authors, namely Witten and Tibshirani~\cite{witten2010framework} and Fulcher et al.~\cite{fulcher2013highly}, propose a couple of alternatives to our feature selection approaches for clustering and respectively time series analysis in general (with applications to time series comparison and clustering).
Although the former authors do not focus on time series explicitly and the latter do not specifically address count time series and value sparseness, we believe those approaches could be used with sparse count time series and in particular to the activity time series we observe.
Hence, we applied Witten and Tibshirani's K-Means and hierarchical-based sparse clustering approaches to our activity-based count time series, thereby taking care to adjust the hyperparameters to our data. We report the results of the application of their method in comparison to our own later on.
Fulcher et al.'s approach, however, would result in the same feature filtering approach we outlined above, as their proposed features include much of the information-theoretic, model-based and locality-focused features we explicitly excluded from our analysis. 
To sum up, the features we find lead to clearly separated, interpretable clusters, as we see in the upcoming section.

\subsection{Clustering Process}

The combination of these three features we propose allows us to derive cohesive, well-separated and interpretable clusters.
Each one of the binary features partitions the space of activity-based time series in two sets.
Those two features thus yield, when combined, four clusters, since they do not capture the same properties of activity-based time series.
The third feature, ratio of unique non-zero values, is continuous and takes values in the interval $[0, 1]$.
This continuous feature measures more granular variations in activity frequency than those afforded by having just the two binary features.
Using this continuous feature by itself, however, does not separate the space in clusters.

We employ the commonly-used unsupervised clustering algorithm K-Means \cite{lloyd1982least}, with k-means++ cluster center initialization~\cite{arthur2007k}, to group similarly active users.
We measure time series similarity with the Euclidean distance on the extracted features.
We briefly explain K-Means: The algorithm begins with a random initialization of K cluster centers, so-called centroids, as K randomly chosen vectors from an input space. The algorithm labels input vectors with the centroid most similar to each of them. Then, it reassigns all K cluster centroids to each cluster's mean vector. These two steps are repeated until convergence \cite{lloyd1982least}. 
We also experimented with both variations of K-Means such as bisecting K-Means~\cite{steinbach2000comparison} as well as with other clustering algorithms such as Ward hierarchical clustering~\cite{ward1963hierarchical} and DBSCAN~\cite{ester1996density}, but those efforts yielded similar results, as we see below.

\noindent \textbf{Selecting the number of clusters.}
The main hyperparameter of K-Means is $K$, representing the number of clusters, which is often a function of expert knowledge or other external factors. However, we aim to learn a suitable number of clusters directly from the data. Therefore, we automate the estimation of K.
The \textit{elbow method}~\cite{ketchen1996application} executes K-Means clustering for a range of values of K and stores the mean distance of centroids to the clustered input, which is termed the cost function, for each K.
With the elbow method, one then graphically identifies the optimal K as the value $K^*$ where the cost function, plotted as a function of K, results in the best trade-off between low cost and maximum cost reduction with respect to $K^*-1$'s cost.
Intuitively, this description of K* matches the point where the cost function forms an ``elbow'', hence the method's name.

We employ a purely numeric method to choose the value for K, since we aim to automatize the search for $K^*$ for a large number of time series.
Similarly to the elbow method, we estimate a statistic on the quality of the clustering for a range of values of K.
Thus, we pick the value $K^*$ that maximizes the \textit{silhouette coefficient}~\cite{rousseeuw1987silhouettes, kaufman2009finding}, which combines statistics on the cluster cohesion (intra-cluster) and separation (inter-cluster) into a single value.
Cluster cohesion, represented by $a_i$, captures the mean distance of an element $i$ in a cluster to other elements in the same cluster.
Cluster separation, represented by $b_i$, denotes the mean distance of an element $i$ in a cluster to other elements in the closest neighboring cluster.
These two factors form the equation for the silhouette coefficient $s_i = (b_i - a_i)/max(a_i, b_i)$, where $-1 \leq s_i \leq 1$.
A high silhouette coefficient implies that the cluster distance of $i$ to other elements in its cluster is low, relative to the mean distance to elements in the next nearest cluster, suggesting the correct assignment of $i$.
The opposite holds for low silhouette coefficient values.

With the application of K-Means for $K = 2, \ldots, 10$ on the extracted features, we look for $K^*$.
We validate separation and cohesion of the $K^*$ clusters graphically with PCA projections into a two-dimensional space (cf. Figure~\ref{fig:silhouette_maximization}).
To check the validity of the clustering obtained with K-Means, we compare its performance with a random clustering baseline, which randomly assigns each input vector to one of $K$ clusters.
Furthermore, as previously mentioned, we compared K-Means with other clustering algorithms: bisecting K-Means, Ward hierarchical clustering and DBSCAN.

\noindent \textbf{Measuring clustering performance.} To measure the clustering performance, we first perform random clustering as a baseline. The random clustering yields $K^*=2$ with average silhouette coefficient values in the interval $[-0.05, 0.02]$.
We then cluster activity-based time series of our datasets and obtain significantly better results.
For all $50$ Stack Exchange instances we obtain average silhouette coefficient values of at least $0.9$ for $K^*$.
$K^*=4$ for $39$ of our $50$ Stack Exchange instances. The remaining $11$ Stack Exchange instances feature a strictly higher optimal number of clusters between six and ten.

Our experiments with other clustering approaches yielded very similar results: Bisecting K-Means and Ward hierarchical clustering return the same average silhouette coefficient values up to a factor of $10^{-3}$ and agree on $K^*$ for all $50$ datasets.
DBSCAN, however, had lower average silhouette coefficient values of at least $0.89$, but also yielded $K^* = 4$ for the same $39$ datasets as before. However, on those $11$ Stack Exchange instances with $K^*>4$, disagreement in both average silhouette coefficient values as well as $K^*$ was highest in the comparison with the other clustering algorithms.

We attribute the similarity of results for different clustering algorithms to the two binary features strongly influencing the distribution of user activity-based time series features in the three-dimensional feature space.
We observe lower silhouette values and disagreement in $K^*$ between the clustering algorithms in cases where the continuous feature plays a larger role in the distribution of a Stack Exchange instance's user activity-based time series features.
DBSCAN seems most sensitive to these feature distributional changes, as its density region-based clustering approach consistently groups points with different binary feature values but similar continuous feature values in one cluster. This clustering behavior, in turn, leads to lower silhouette and lower $K^*$ than other clustering algorithms agree upon.
We stress that other binary features might have lead to same high silhouette coefficient results, but they lead to ultimately different results and interpretation of user behavior.

The best results we achieved with the K-Means and hierarchical sparse clustering approaches by Witten and Tibshirani yielded $K^* = 2$ and silhouette coefficient values of a maximum of $0.88$ and significantly lower for all $K > 2$.
We believe tailoring these algorithms to find more granular structure in sparse count time series data such as ours to be an interesting avenue of future work.

Finally, in two-dimensional projections of the clusters with PCA, we observe clear graphical separation for the $K^*$ clusters in most Stack Exchange instances.

\subsection{Analyzing Cluster Properties}
We analyze the clusters we obtain to better understand the activity composition captured by K-Means. To that end, we start by computing basic descriptive statistics on the clusters, such as their size, as measured by the number of users per cluster. Further, we plot the activity-based time series closest to each centroid and thereby visualize typical activity profiles for each cluster.
We then visually inspect the sum of the activities in each of the clusters to discern overall cluster group dynamics.
We corroborate this visual inspection with a quantification of the relative sizes of the clusters as the fraction of a cluster's activity in total activity.
Finally, we look for commonalities in these patterns between Stack Exchange instances, and assess and discuss their practical relevance in for Q\&A community building efforts.

\section{Results}
\label{sec:results}

\begin{figure}
    \begin{subfigure}[t]{.24\textwidth}
        \centering
        \includegraphics[width=1.0\textwidth]{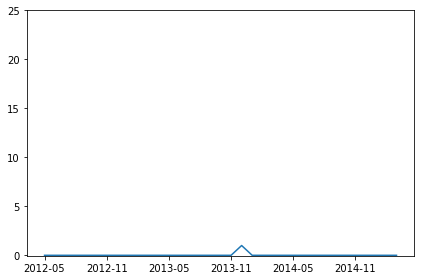}
        \caption{Non-Recurring}
        \label{fig:archetypes_singular}
    \end{subfigure}
    \begin{subfigure}[t]{.24\textwidth}
        \centering
        \includegraphics[width=1.0\linewidth]{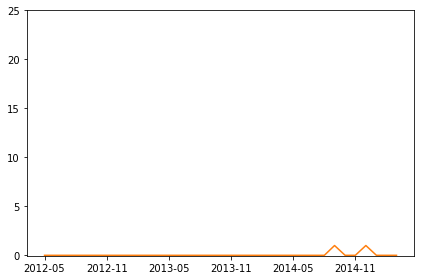}
        \caption{Sporadic}
        \label{fig:archetypes_sporadic}
    \end{subfigure}
    \begin{subfigure}[t]{.24\textwidth}
        \centering
        \includegraphics[width=1.0\linewidth]{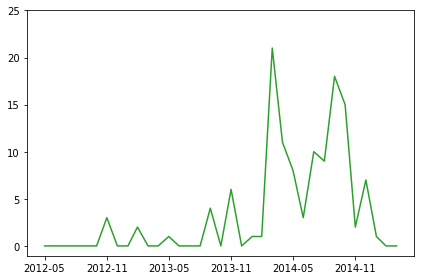}
        \caption{Frequent}
        \label{fig:archetypes_frequent}
    \end{subfigure}
    \begin{subfigure}[t]{.24\textwidth}
        \centering
        \includegraphics[width=1.0\linewidth]{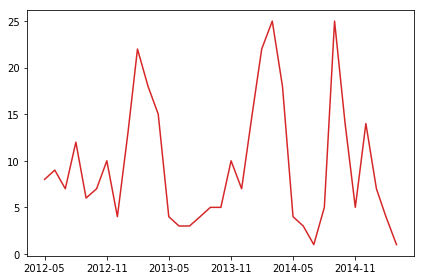}
        \caption{Permanent}
        \label{fig:archetypes_sustained}
    \end{subfigure}
    \caption{\label{fig:archetypes}
    \textbf{Activity Archetypes.} We illustrate typical profiles of activity-based time series nearest to K-Means centroid for $K^*=4$.
    Users of the \aanonrecurring (a) often feature one single, isolated peak of activity.
    Users of the \aasporadic (b) typically exhibit a few isolated activity peaks of equal height.
    Users of the \aafrequent (c) show varying but regular activity over time.
    Finally, repeatedly high levels of activity over time characterize users of the \aapermanent (d). 
    In short, we observe that user activity can be grouped into these four activity profiles, which mainly capture different degrees of frequency in user activity.
    }
\end{figure}

\subsection{Activity Archetypes}
\label{sec:archetypes}

For all Stack Exchange instances with $K^*=4$, we observe four commonly occurring types of temporal user activity patterns, which we term \aats (see Figure~\ref{fig:archetypes}).
The patterns of these time series are representative of the four \aats, which we describe in ascending order of activity frequency and volume. 

In general, users of the \textbf{\aanonrecurring} (see Figure~\ref{fig:archetypes_singular}) exhibit one prominent peak of activity.
Taking the median value over all Stack Exchange instances, we find typical users of the \aanonrecurring post $1$ question and write $1.27$ answers or comments in a median of $1$ active month (i.e. the number of months in which an user posted at least one question or answer in a given Stack Exchange instance).
Furthermore, their median tenure length, as measured by the difference between their first and last dates of activity (i.e. writing a question or answer), is less than $1$ month.
Users of the \aanonrecurring comprise on average $88.4$\% of total user count of a Stack Exchange instance.
This majority of users thus typically post a question, follows up on it with discussion with the rest of the community in a short, concentrated period of time and does not return.
This suggests users of the \aanonrecurring type have one or two concrete asking needs, which, after some discussion, are satisfied, completing the user's participation in the community.
User ``Amit Kumar Gupta''\footnote{https://psychology.stackexchange.com/users/7338/amit-kumar-gupta} of the Stack Exchange instance \textit{cogsci}\footnote{In December 2017, \textit{cogsci} was renamed to \textit{psychology} (source: https://biology.meta.stackexchange.com/questions/3779/cogsci-has-changed-its-name), but we refer to it by its old name for the sake of consistency with our dataset, which includes data up to February 2017.} and his question on types of memory and ensuing discussion exemplifies this behavior by users of the \aanonrecurring.
For better comparison with other \aats, we use the Stack Exchange instance \textit{cogsci} for further user examples.

The \textbf{\aasporadic} (see Figure~\ref{fig:archetypes_sporadic}) features higher activity levels than the \aanonrecurring.
Users of the \aasporadic write a median of $2.09$ questions and $2.26$ answers or comments in a median of $2.06$ active months.
In contrast to the \aanonrecurring, the median tenure length of the \aasporadic is $6.08$ months and they comprise on average $10.1$\% of total user count of a Stack Exchange instance.
Hence, in comparison with the \aanonrecurring, not only do users of the \aasporadic pose more questions (and answer and discuss them slightly more), but they also do so throughout a remarkably longer period of time. This suggests they lurk and engage more with the Stack Exchange instance community as a whole.
For an example of such user behavior, refer to user ``201044''\footnote{https://psychology.stackexchange.com/users/7340/201044} of \textit{cogsci}.

We observe significantly more activity from users of the \textbf{\aafrequent} (see Figure~\ref{fig:archetypes_frequent}). 
Such users have a median of $19.26$ questions and $28.23$ answers or comments and are active in a median of $12.09$ months out of median tenures of $32.31$ months.
The \aafrequent is notably less numerous, as it accounts for an average $1.3$\% of total user count of a Stack Exchange instance.
We observe a large gap in the activity profile of the \aafrequent and the previous two, as users of the \aafrequent participate in Stack Exchange instance communities with greater quantity and higher frequency. Their remarkably long tenures suggest they accompany community development, despite not being active every month.
Average users of the \aafrequent behave like user ``Greg McNulty''\footnote{https://psychology.stackexchange.com/users/849/greg-mcnulty} of \textit{cogsci}.

The most active group of users we identified belongs to the \textbf{\aapermanent} (see Figure~\ref{fig:archetypes_sustained}). 
As such, this group of users posts a median of $26.12$ questions and $56.68$ answers or comments. They are active in a median of $13.99$ months and their median tenure is $32.86$ months.
On average, users of the \aapermanent represent just $0.2$\% of total user count of a Stack Exchange instance.
Although users of this archetype feature tenures comparable to those of the \aafrequent, the fact they are the most active overall, combined with the fact there are very few of them, could indicate users of the \aapermanent lead activity in the community.
Users such as ``Alex Stone''\footnote{https://psychology.stackexchange.com/users/953/alex-stone} of \textit{cogsci} exemplify and could cement our reading of the role users of the \aapermanent play in Stack Exchange instances.

\noindent \textbf{Feature importance analysis.} To support these descriptions with a quantitative assessment of the four \aats in terms of our three features, we evaluated, separately on the questions and answers of all $50$ Stack Exchange instances, the power of the features in explaining the four \aats with ANOVA~\cite{fox2015applied} and the distribution of the feature's values over the \aats with random forests~\cite{breiman2001random} and in particular also decision trees~\cite{breiman1984classification}.

For the ANOVA approach, we fitted a generalized linear model of the three features per user as independent variables and the \aats resulting from the clustering as dependent variables. As the \aats represent a discrete dependent variable, we assume it is binomially distributed and we use a logit link function.
The ANOVA measure of each feature's effect in such a regression model suggests every feature is significant in explaining the \aats, as the corresponding p-values (for $H_0$: dependent variable's coefficient is 0 tested with an F-test) are all smaller than $8.47 \cdot 10^{-6}$.
These results hold for both questions and answers datasets of each of the $50$ Stack Exchange instances.

\begin{figure}
    \begin{subfigure}[t]{.48\textwidth}
        \centering
        \includegraphics[width=1.0\linewidth]{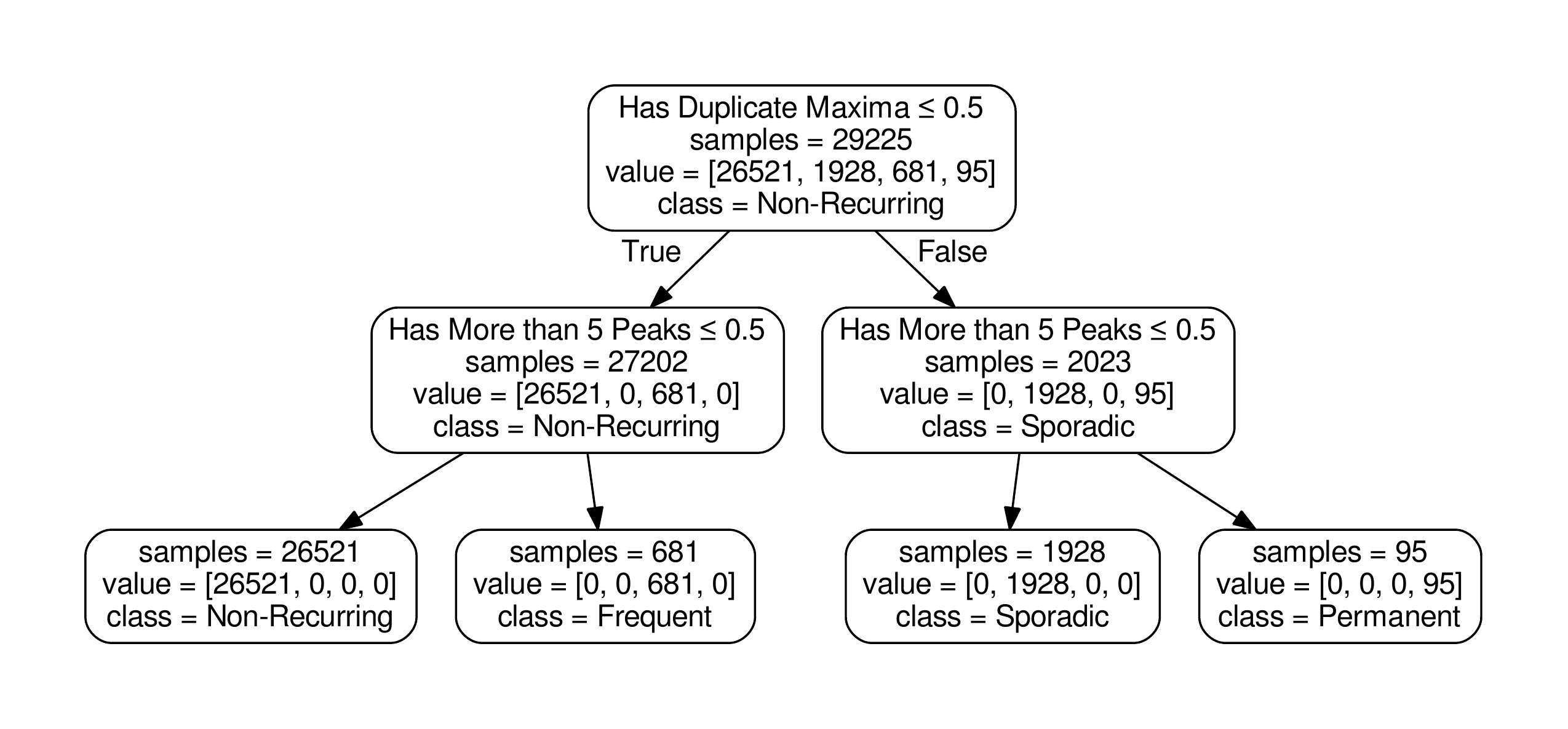}
        \caption{English Answers Decision Tree}
        \label{fig:english_answers}
    \end{subfigure}
    \begin{subfigure}[t]{.48\textwidth}
        \centering
        \includegraphics[width=1.0\textwidth]{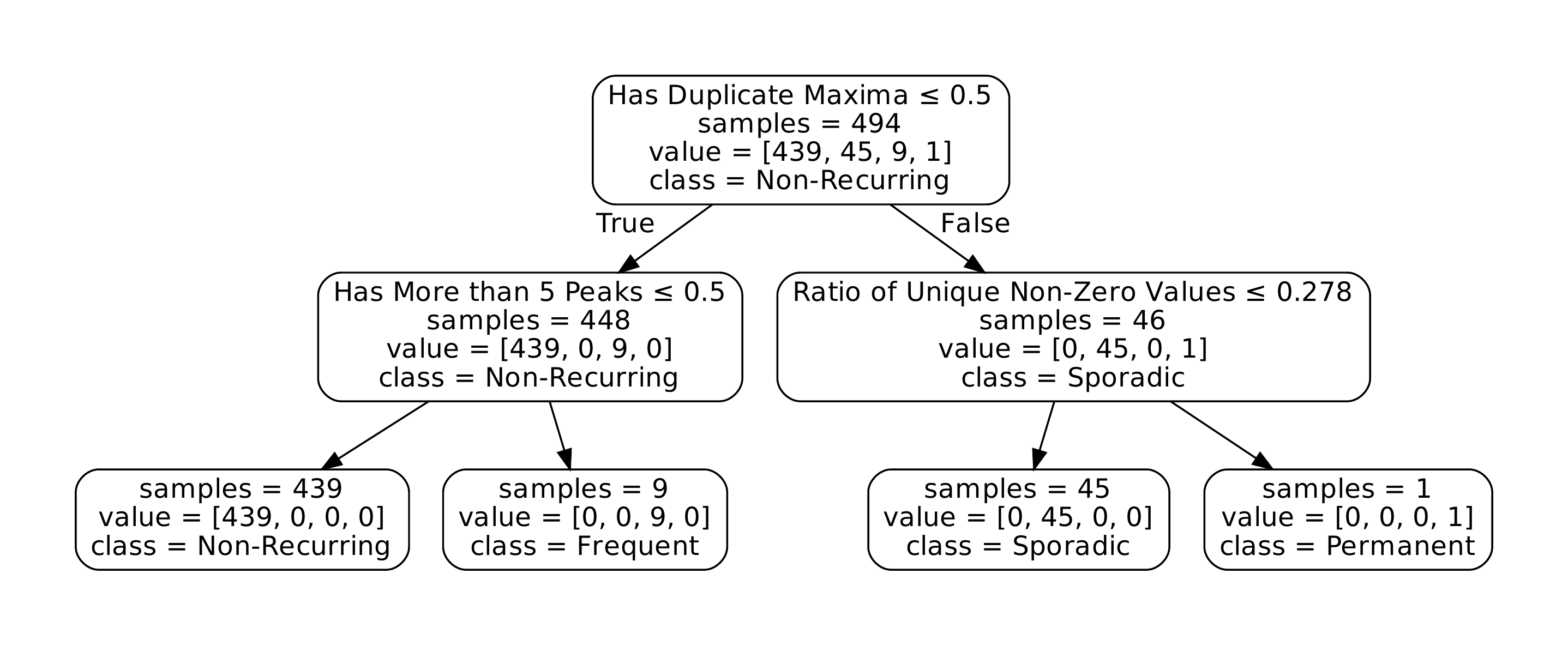}
        \caption{Sustainability Answers Decision Tree}
        \label{fig:sustainability_answers}
    \end{subfigure}
    \caption{\label{fig:decisiontrees}
    \textbf{Decision trees fitted to user activity-based time series features.} We depict the result of applying decision trees to fit the extracted user answer activity-based time series clusters as a function of the three features we propose.
    The pictures show the number of users (samples) per decision tree node out of all clusters, i.e. \aats.
    The decision tree for the \textit{English} Stack Exchange instance (a) shows the defining feature values per \aats, which are dominated by the two binary features, ``duplicate maxima'' and ``more than 5 peaks''.
    The \textit{Sustainability} Stack Exchange instance (b) features the \aats as a function of all three features we presented, as there is enough user behavior variability in \textit{Sustainability} for the continuous feature to offset the dominance by the two binary features.
    Hence, all three features are important to characterize user behavior, but their importance varies with the Stack Exchange instance.
    }
\end{figure}

In a similar experiment, we fitted random forests on the three user features over all $50$ Stack Exchange instances (again, separately for questions and for replies) to explain the \aats.
One of the outputs of random forests is estimated feature importance.
In that regard, the random forests' output agrees with ANOVA's: All three features are important to classify \aats in both their questions and answers activity.
Moreover, random forests output a numeric estimation of feature importance for classification on a scale from zero to one: for questions-based (answers-based) activity, $0.695$ ($0.187$) is the feature importance of the ratio of unique non-zero values to time series length, $0.271$ ($0.614$) the one of the feature capturing if the time series has more than five peaks and $0.034$ ($0.199$) the one of the feature regarding duplicate maxima.
    
Breaking down this high-level view of all $50$ Stack Exchange instances by instance allows us to visualize the feature values composing each of the \aats and clusters we find. 
To that end, we visualize, in Figure~\ref{fig:decisiontrees}, a decision tree fitted to answers-based activity of the \textit{English} (Figure~\ref{fig:english_answers}) and \textit{Sustainability} (Figure~\ref{fig:sustainability_answers}) Stack Exchange instances.
Furthermore, we show the number of users at each node (in total, as given by ``samples'' and per class, as given by ``values'') in the decision tree's path, and the resulting class, i.e. Activity Archetype.
We observe \textit{English} clearly distinguishes the four \aats along the values of the two Boolean features ``duplicate maxima'' and ``more than 5 peaks''. 
The decision tree for the Stack Exchange instance \textit{Sustainability} makes use of the two Boolean features, as well as the continuous feature ``ratio of unique non-zero values'', to classify the four \aats.
We relate this fact to this Stack Exchange instance already including \aats, but with slightly more variability in them (as captured by the continuous feature).
    
Note, however, that not all Stack Exchange instances feature such temporal user activity patterns as given by the four \aats.
The structure of the decision trees of such instances included more levels and a number of nodes on the feature ``ratio of non-zero unique values''.
When $K^*>4$, the temporal user activity patterns we observe represent more granular variations of the four \aats we highlight, as exemplified in the legend of Figure~\ref{fig:answers_transient}.
As we find more than ten different variations of this kind, we do not characterize them in more detail.

\begin{figure}[!t]
	\begin{subfigure}{.48\textwidth}
		\centering
		\includegraphics[width=6.3cm,height=6.3cm,keepaspectratio]{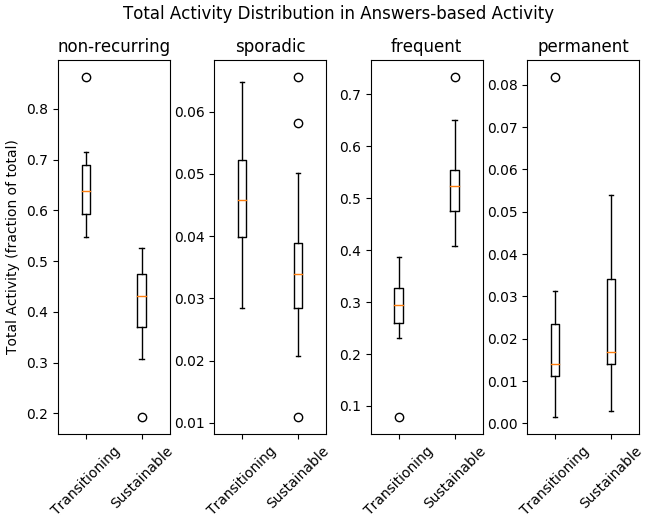}
		\caption{Total activity distribution of answers-based activity time series}
		\label{fig:totalactivity_answers}
	\end{subfigure}
	\begin{subfigure}{.48\textwidth}
		\centering
		\includegraphics[width=6.3cm,height=6.3cm,keepaspectratio]{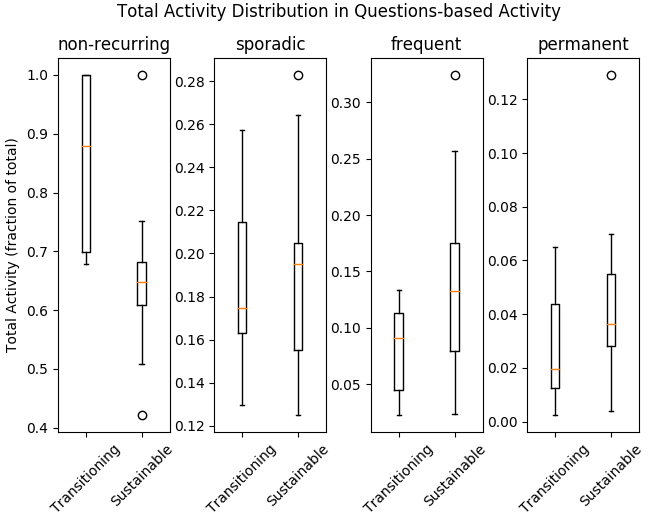}
		\caption{Total activity distribution of questions-based activity time series}
		\label{fig:totalactivity_questions}
	\end{subfigure}
	\caption{\label{fig:totalactivity_boxAndWhiskers}
		\textbf{Distinction between \aatransitioning and \aasustainable Stack Exchange instances.}
		For all Stack Exchange instances of the types \aatransitioning and \aasustainable, we depict the fractions of total answers-based (Fig.~\ref{fig:totalactivity_answers}) and questions-based (Fig. \ref{fig:totalactivity_questions}) activity generated per \textit{Activity Archetype}. 
		In \aatransitioning (\aasustainable) instances, answer-based activity is dominated by \aanonrecurring (\aafrequent) users, which contribute a median fraction of $0.63$ ($0.52$) of total answer activity.
		Overall, we observe stark contrasts in contributions of \aats to total activity of different Stack Exchange instances.
	}
\end{figure}

\subsection{Composition of Stack Exchange Instances}
A total of $39$ Stack Exchange instances exhibit $K^*=4$, i.e. the four \aats, in our clustering experiment.
First, we categorize these Stack Exchange instances with a breakdown of their total question and answer-based activity by \aats.
We call this breakdown of activity by \aats the \aaac of a Stack Exchange instance and we find two types of \aaac.
Then, we analyze how the \aaac changes over time.

\textbf{Derivation and analysis of the \aaac of Stack Exchange instances.}
Among those $39$ instances, we observe two distinct \aaacs with respect to the contribution to total answer activity by users of the \aanonrecurring.
Recall users of the \aanonrecurring represent the majority at an average fraction of $88.4$\% of total user count.
Interestingly, in some instances, they do not account for the majority fraction of total answer activity and play a less prominent role in total question activity.
However, as might be expected, users of the \aanonrecurring dominate the \aaac in other instances.

Therefore, we derive two distinct groups of Stack Exchange instances by setting the following threshold: 
In a given Stack Exchange instance, if users of all \aats except for the \aanonrecurring account for $90\%$ or more of answer-based activity by the \aanonrecurring, we categorize the instance as \aasustainable, otherwise as \aatransitioning. We experimented with variations of the $90\%$ threshold in the range $[85\%, 95\%]$, but we did not arrive at remarkably different results and conclusions.
Recall our dataset includes, besides the $39$ Stack Exchange instances with $K^*=4$, a total of $11$ instances with $K^*>4$.
We name this group of instances \aaemerging, but, for now, we focus on \aasustainable and \aatransitioning instances.
As we discuss later, this naming choice correlates with key developmental characteristics of the two types of Stack Exchange instances.
Using this criterion, we identify $26$ \aasustainable Stack Exchange instances (of which $5$ still are in the Area 51 incubator)\footnote{The 26 \aasustainable Stack Exchange instances are \textit{english}, \textit{unix}, \textit{softwareengineering}, \textit{gaming}, \textit{tex}, \textit{stats}, \textit{wordpress}, \textit{physics}, \textit{mathoverflow}, \textit{sharepoint}, \textit{scifi}, \textit{ux}, \textit{webmasters}, \textit{graphicdesign}, \textit{workplace}, \textit{salesforce}, \textit{cs}, \textit{bicycles}, \textit{skeptics}, \textit{christianity}, \textit{sound}, \textit{history}, \textit{gardening}, \textit{linguistics}, \textit{outdoors} and \textit{tridion}, with \textit{history}, \textit{gardening}, \textit{linguistics}, \textit{outdoors} and \textit{tridion} still being in the Area 51 incubator as of 02/13/2017.} and $13$ \aatransitioning Stack Exchange instances (with $8$ of them still in the Area 51 incubator)\footnote{The \aatransitioning group of Stack Exchange instances consists of \textit{bitcoin}, \textit{chemistry}, \textit{chess}, \textit{codereview}, \textit{cogsci}, \textit{music}, \textit{opendata}, \textit{philosophy}, \textit{poker}, \textit{reverseengineering}, \textit{space}, \textit{sports} and \textit{sustainability}. As of 02/13/2017 \textit{chemistry}, \textit{codereview}, \textit{music} and \textit{philosophy} have left the Area 51 incubator.}.

We compare the \aaac of the \aatransitioning and \aasustainable Stack Exchange instance types in more detail in Figure~\ref{fig:totalactivity_boxAndWhiskers}.
We observe the highest proportion of answers-based activity in \aasustainable Stack Exchange instances comes from the \aafrequent, whereas the \aanonrecurring generates most (questions and) answers-based activity in \aatransitioning Stack Exchange instances.
We draw these conclusions from the relatively higher (lower) median total activity fraction values for users in the \aafrequent (\aanonrecurring) in \aasustainable instances compared to \aatransitioning instances (Figure~\ref{fig:totalactivity_answers}). 
Furthermore, we highlight the relative importance of the \aanonrecurring and the \aasporadic in questions-based activity (see Figure~\ref{fig:totalactivity_questions}): Although more so in \aatransitioning Stack Exchange instances, both still play a significant role in the \aasustainable instance type.
Differences between instance types in the role of the \aapermanent are qualitatively the same as in the \aafrequent but to a lesser degree, as the \aapermanent accounts for a median fraction of only $0.014$ (respectively $0.018$) of total questions and $0.02$ ($0.037$) of total answers in the \aatransitioning (\aasustainable) instance types.

\begin{figure*}
    \begin{subfigure}{.32\textwidth}
        \centering
        \includegraphics[width=1.0\linewidth]{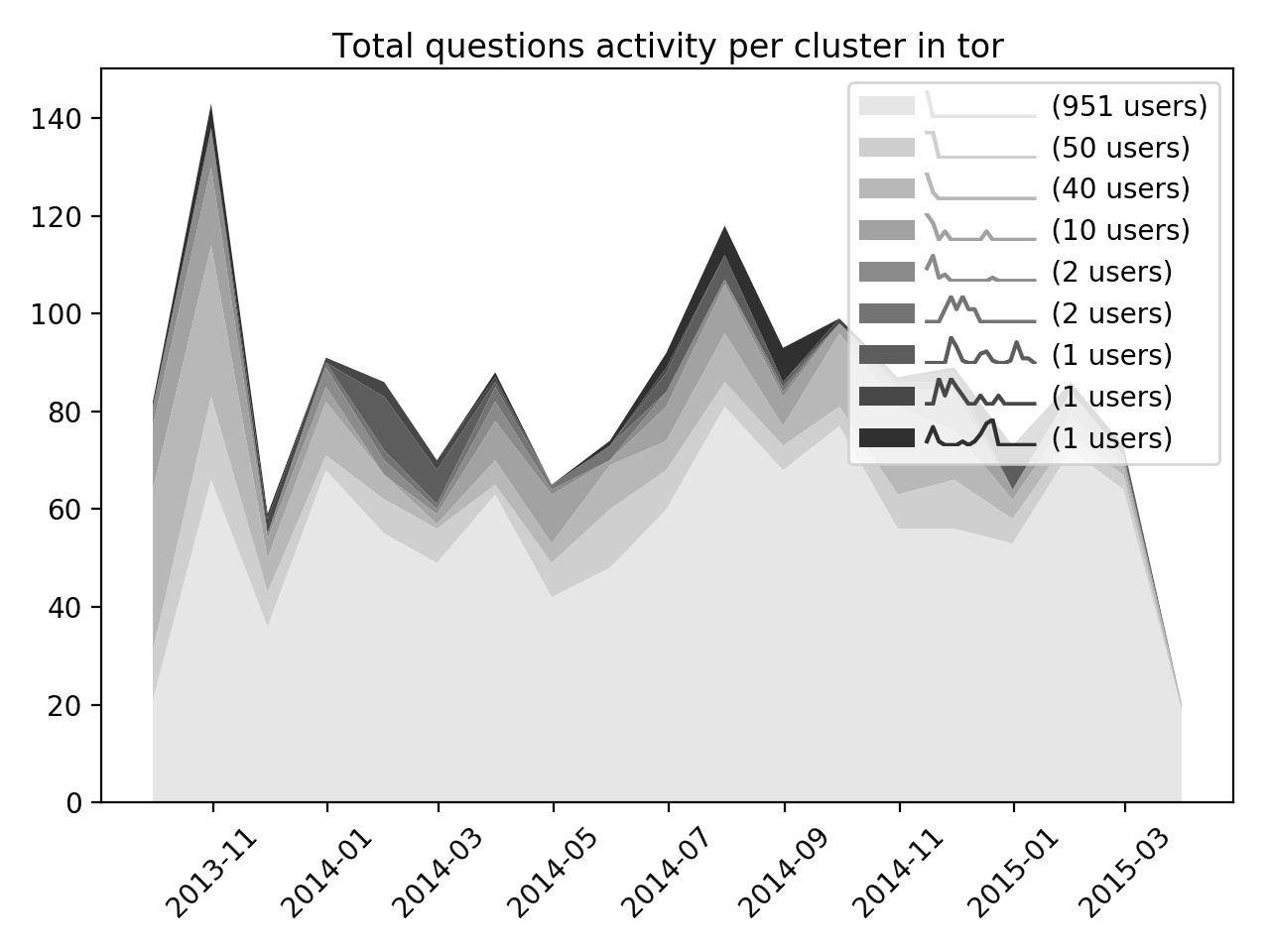}
        \label{fig:questions_transient}
    \end{subfigure}
    \begin{subfigure}{.32\textwidth}
        \centering
        \includegraphics[width=1.0\linewidth]{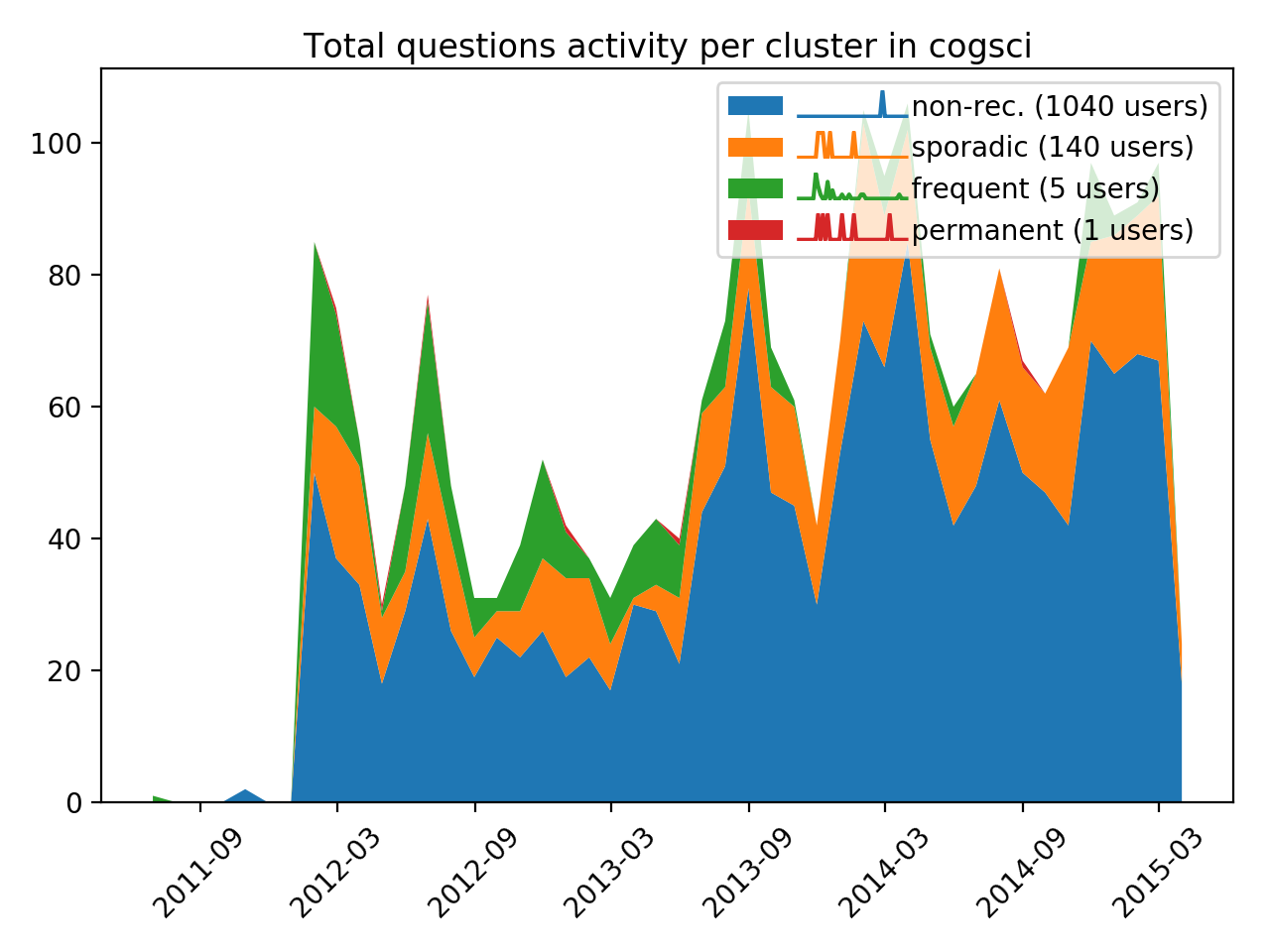}
        \label{fig:questions_shifting}
    \end{subfigure}
	\begin{subfigure}{.32\textwidth}
        \centering
        \includegraphics[width=1.0\linewidth]{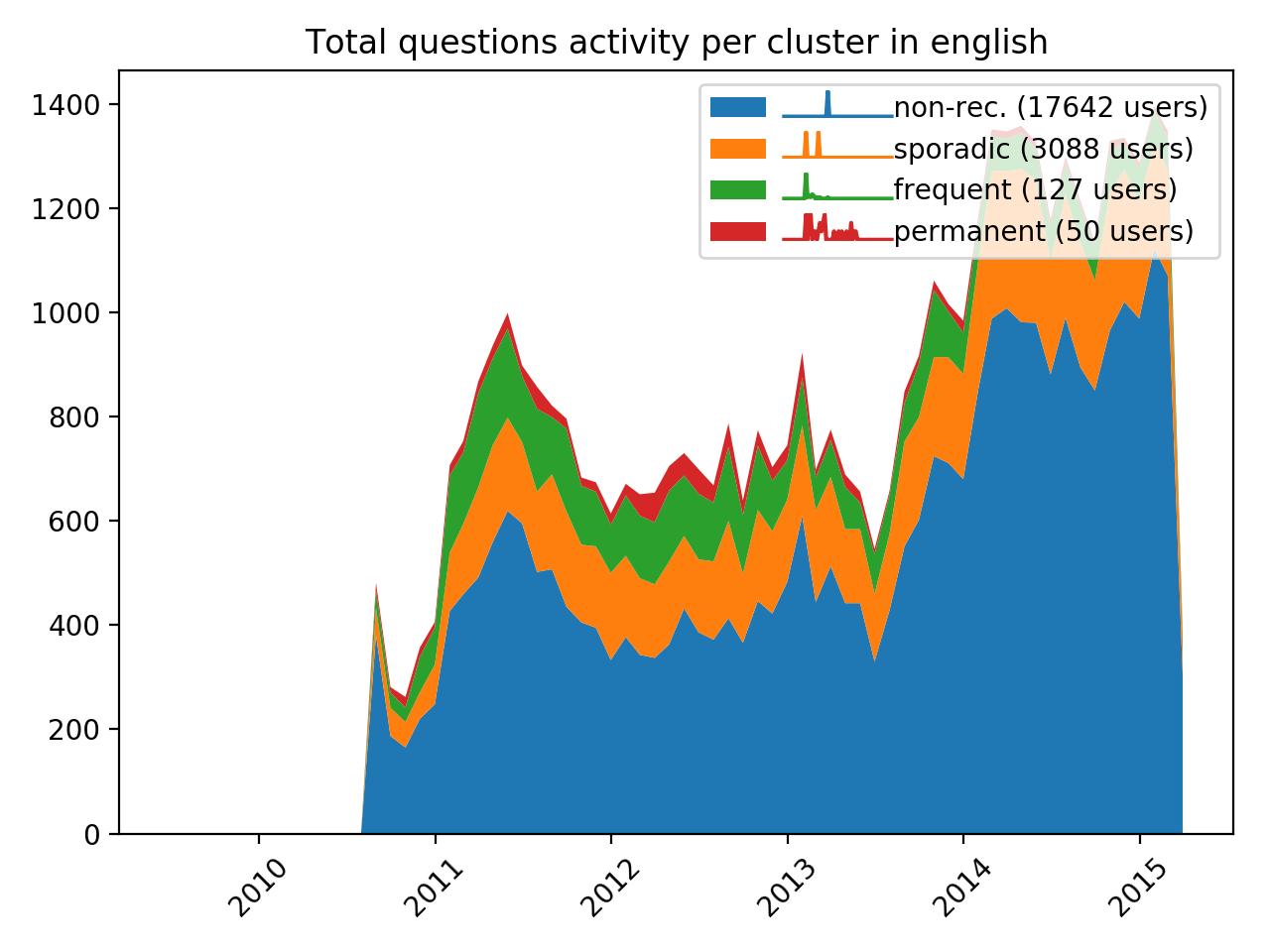}
        \label{fig:questions_stable}
    \end{subfigure}\\[-3ex]
    \begin{subfigure}{.32\textwidth}
        \centering
        \includegraphics[width=1.0\linewidth]{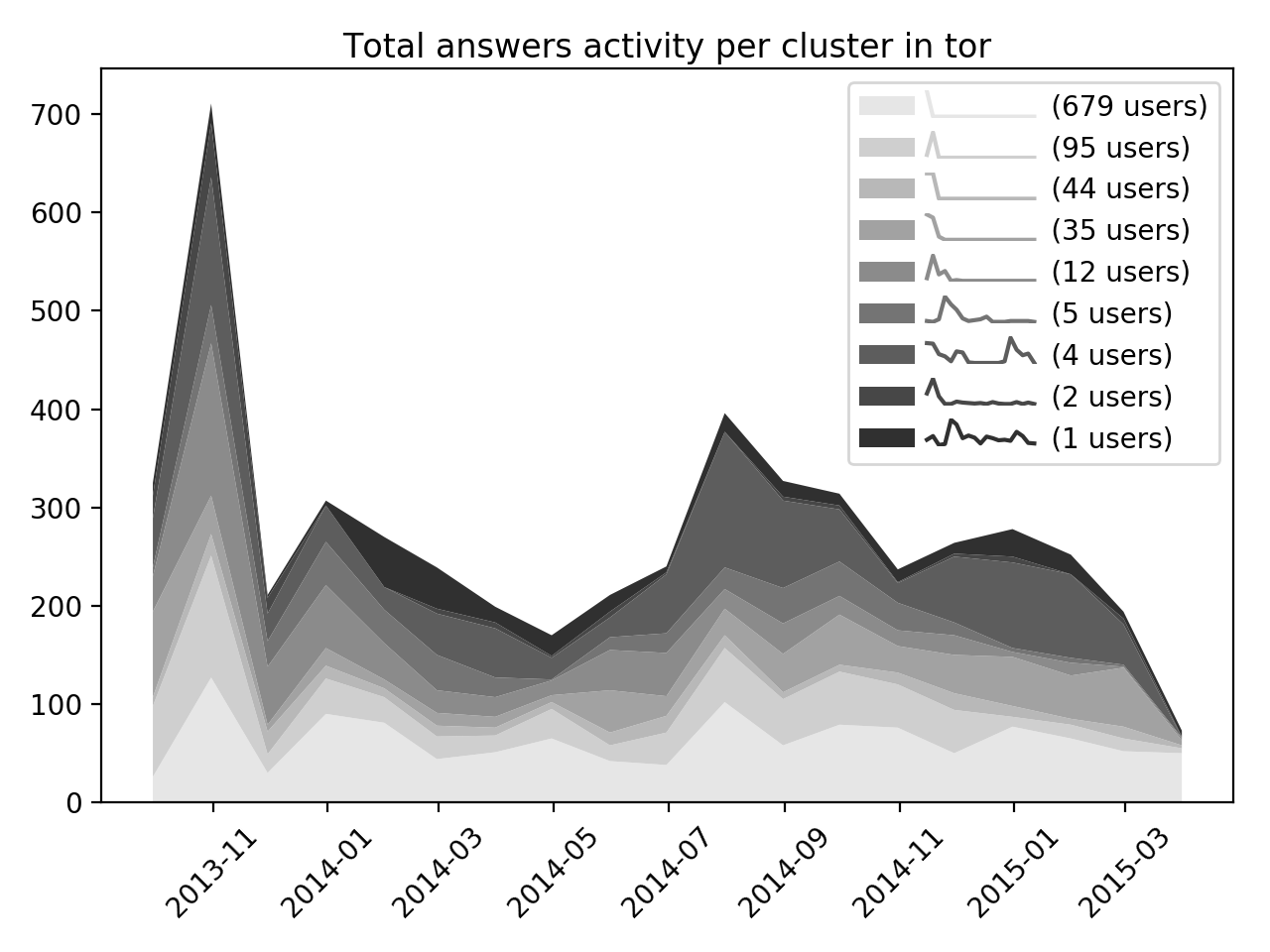}
        \caption{\aaemerging}
        \label{fig:answers_transient}
    \end{subfigure}
    \begin{subfigure}{.32\textwidth}
        \centering
        \includegraphics[width=1.0\linewidth]{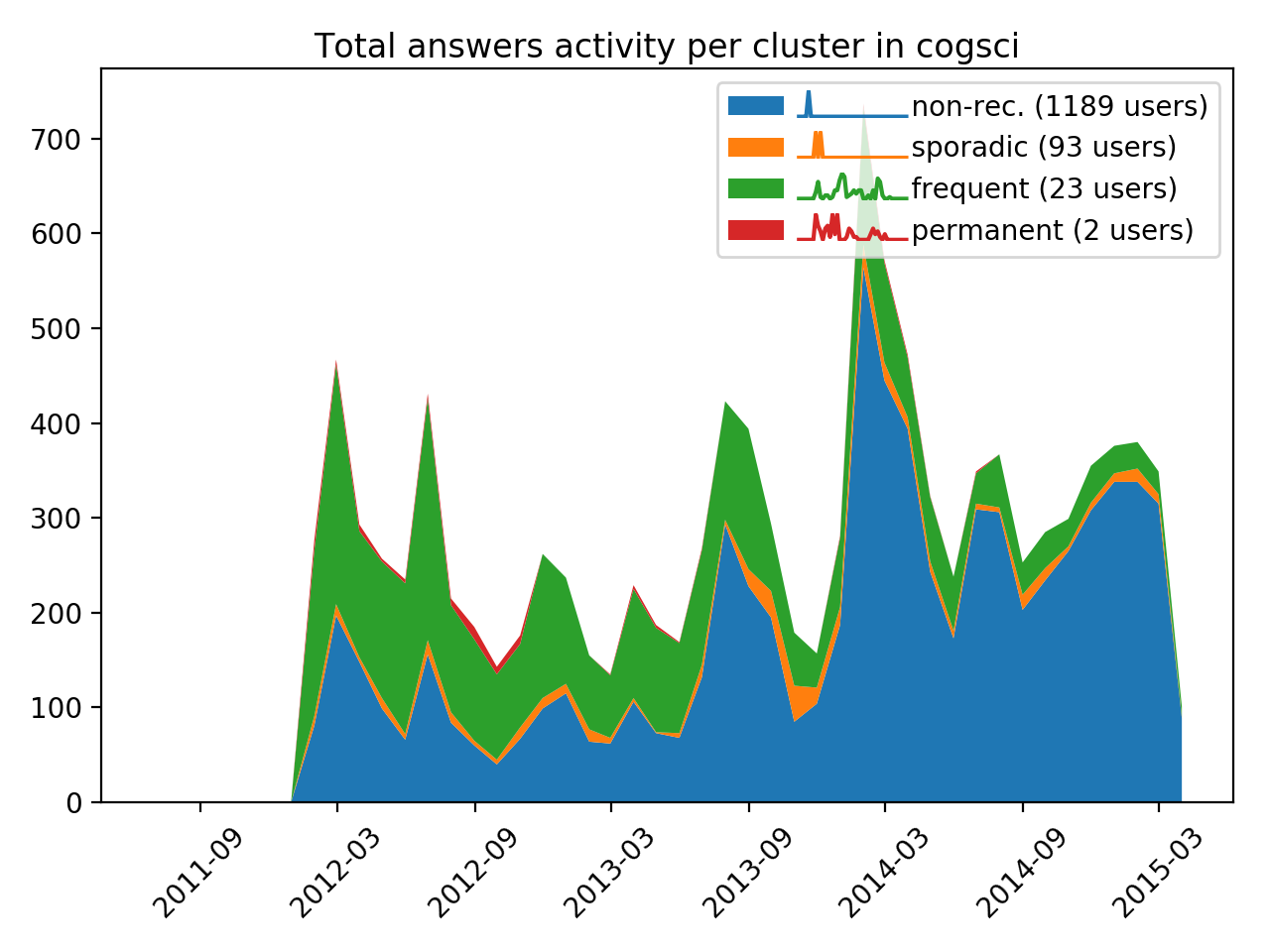}
        \caption{\aatransitioning}
        \label{fig:answers_shifting}
    \end{subfigure}
    \begin{subfigure}{.32\textwidth}
        \centering
        \includegraphics[width=1.0\linewidth]{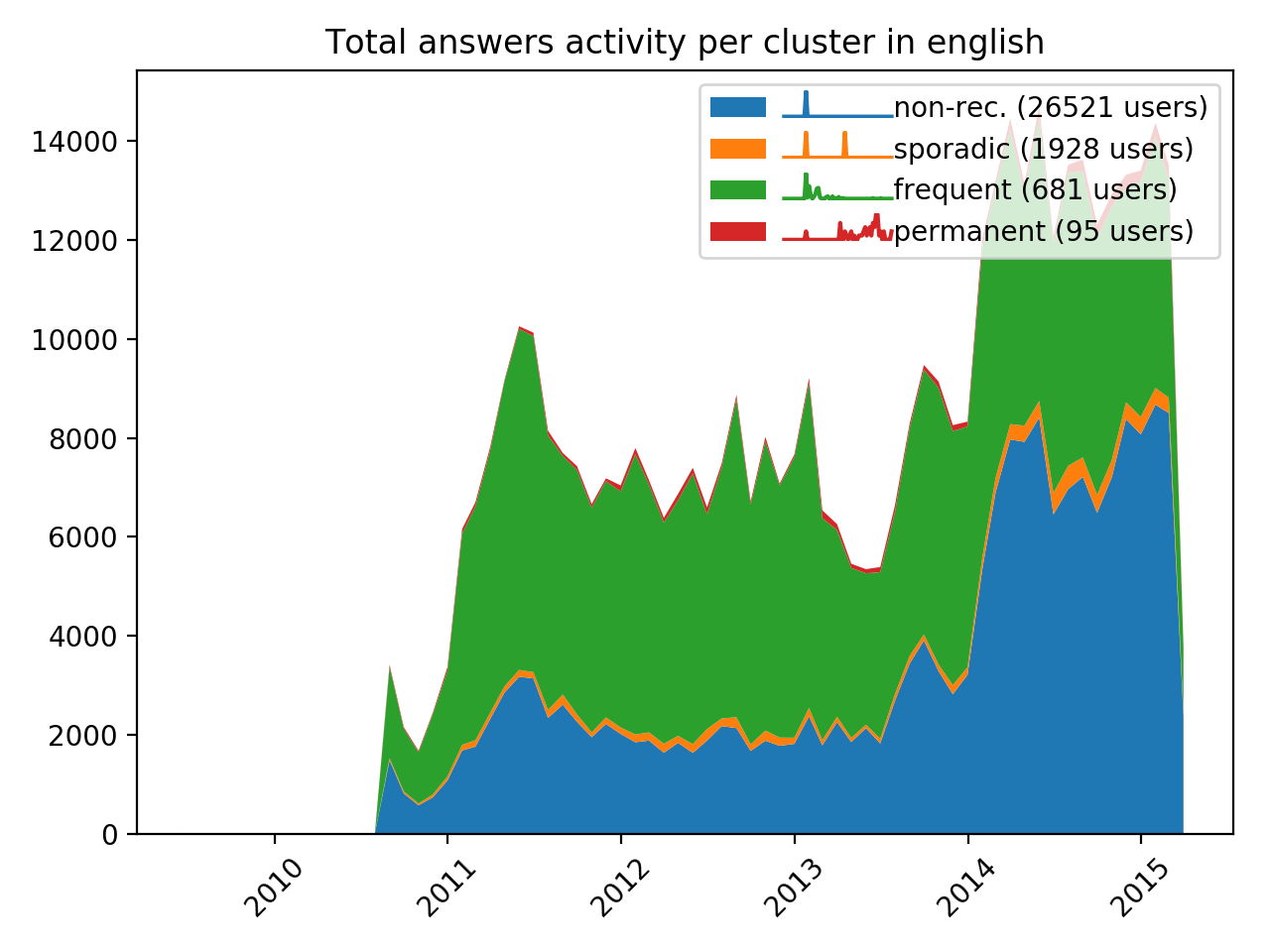}
        \caption{\aasustainable}
        \label{fig:answers_stable}
    \end{subfigure}   
    \caption{\label{fig:activityPerCluster}
        \textbf{Temporal dynamics of the three types of \aaac of Stack Exchange instances: \aaemerging, \aatransitioning and \aasustainable.}
        We plot the total count of questions-based (top) and answers-based (bottom) activity of three Stack Exchange instances over time, and break these activity totals down by \aat.
        A high value of $K^*$, indicating the four user archetypes do not prevail, characterizes \aaemerging instances like \textit{tor} (Fig.~\ref{fig:answers_transient}).
        Notably, we observe similarly low levels of activity in \textit{tor} and \aatransitioning instances like \textit{cogsci} (Fig.~\ref{fig:answers_shifting}), with \textit{tor} having overall declining activity and \textit{cogsci} oscillating around positive growth.
        The \aasustainable instance \textit{English} (Fig.~\ref{fig:answers_stable}), however, exhibits high activity levels, and pronounced growth in activity.
        These aspects hint at a link between \aaac and overall activity development.
    }
\end{figure*}

\textbf{Contextualization of the \aaac of Stack Exchange instances.}
We feature a graphical comparison of total activity volume and the \aaac in instances representative of the three instance types in Figure~\ref{fig:activityPerCluster}.
We draw a connection between the \aaac and key developmental statistics of the instances as summarized in Table~\ref{table:categorized_instance_statistics}.

We address the \textbf{\aaemerging} group of Stack Exchange instances\footnote{The \aaemerging group of Stack Exchange instances consists of \textit{arduino}, \textit{buddhism}, \textit{earthscience}, \textit{ebooks}, \textit{freelancing}, \textit{ham}, \textit{joomla}, \textit{lifehacks}, \textit{puzzling}, \textit{startups} and \textit{tor}. Only \textit{puzzling} has left the Area 51 incubator as of 13. February 2017.} first.
\aaemerging Stack Exchange instances do not exhibit the \aats defined in Section~\ref{sec:archetypes}, but instead feature more variations thereof.
In general, \aaemerging instances are among the newest, least active and smallest out of the $50$ instances we consider:
three out of five smallest instances listed in Table~\ref{table:categorized_instance_statistics} belong to the \aaemerging group.
Moreover, these instances feature an overall negative activity growth\footnote{Note that we estimate activity growth as the slope of a linear regression on total activity (dependent variable) per month and year (independent variable) fitted with ordinary-least-squares and normalized with a min-max transformation for comparing instances (see Table~\ref{table:categorized_instance_statistics}).}, i.e. these instances' activity levels drop on average.
Furthermore, ten out of eleven \aaemerging Stack Exchange instances are still in the Area 51 incubator.
Figure~\ref{fig:answers_transient} illustrates a typical activity profile of \aaemerging instances, as exemplified by the Stack Exchange instance \textit{tor}.

As previously discussed, in \textbf{\aatransitioning} Stack Exchange instances, users of the \textit{Non-Recurring} and \textit{Sporadic Activity Archetypes} generate the most activity, with the \aasporadic acting more prominently in questions-based than answers-based activity.
The activity dynamics of \aatransitioning Stack Exchange instances exhibit strong oscillations over time, as exemplified by the Stack Exchange instance \textit{cogsci} in Figure~\ref{fig:answers_shifting}.
We note that some of the \aatransitioning Stack Exchange instances are among the five smallest datasets in our analysis, as Table~\ref{table:categorized_instance_statistics} indicates.
Other \aatransitioning instances vary considerably in numbers of users and age, and the Stack Exchange instance \textit{codereview} has one of the largest user bases with $19,140$ users and features very high activity levels at a total of $157,593$ questions and answers.
Overall, however, the average activity growth of all \aatransitioning instances is about $0$. In other words, these instances' activity levels oscillate (and stagnate) over the course of their existence.

\begin{table}[b]
    \tiny
    \centering
    \caption{\label{table:categorized_instance_statistics} 
    \textbf{Statistics on largest and smallest Stack Exchange instances.} For the top and bottom five Stack Exchange instances with most and respectively least users, we list a number of statistics, sorted by the number of users:
    Instance type, number of users, total activity (i.e. sum of questions and answers), age in months and the slope of the trend of total activity (dependent variable) per month and year (independent variables).
    The top five Stack Exchange instances are all of the \aasustainable type, and feature a positive growth trend.
    In contrast to those instances, the bottom five Stack Exchange instances are either \aaemerging or \aatransitioning and have dwindling growths (negative trend slope).
    }
    \begin{threeparttable}
    \begin{tabular}{l|c|c|c|c|c}
        \toprule
        \textbf{Instance name}&\textbf{Instance type}&\textbf{Users}&\textbf{Activity}&\textbf{Months}&\textbf{Trend slope}\\ \midrule
        english&\aasustainable&$37125$&$522128$&$70$&$0.013$\\
        unix&\aasustainable&$36397$&$390930$&$80$&$0.012$\\
        softwareengineering&\aasustainable&$35816$&$467234$&$80$&$0.006$\\
        gaming&\aasustainable&$34641$&$321857$&$68$&$0.007$\\
        tex&\aasustainable&$31039$&$624166$&$80$&$0.014$\\
        \midrule
        poker&\aatransitioning&$594$&$5185$&$39$&$-0.002$\\
        earthscience&\aaemerging&$578$&$5981$&$12$&$-0.040$\\
        sustainability&\aatransitioning&$555$&$5274$&$27$&$-0.015$\\
        ebooks&\aaemerging&$501$&$3094$&$16$&$-0.041$\\
        ham&\aaemerging&$473$&$5023$&$18$&$-0.037$\\
        \bottomrule
    \end{tabular}
    \end{threeparttable}
\end{table}

On the other hand, in \textbf{\aasustainable} Stack Exchange instances such as \textit{english}, users of the \aafrequent generate the most answers-based activity, despite, again, representing a reduced percentage of total user base.
In general, \aasustainable Stack Exchange instances are among the oldest, most active ones, feature with the highest number os users (cf. Table~\ref{table:categorized_instance_statistics}), and exhibit high activity levels and a steady growth of activity (cf. Figure~\ref{fig:answers_stable}).
Furthermore, average activity growth of all \aasustainable instances is positive.

\textbf{Instance type evolution over time.}
\begin{figure*}
    \centering
    \includegraphics[width=11cm,height=4.5cm,keepaspectratio]{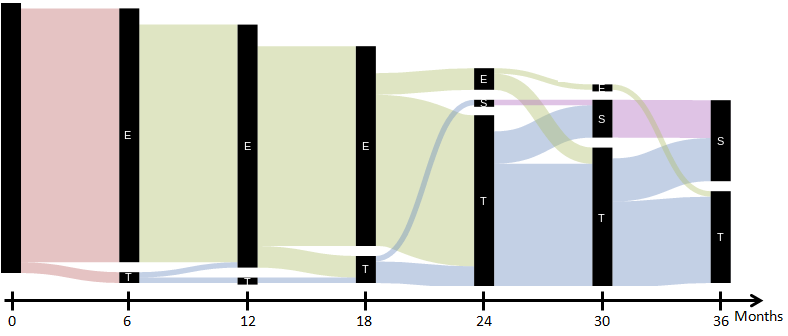}
    \caption{\label{fig:dataset_type_per_quarter}
    \textbf{Temporal evolution of Stack Exchange instance types.} 
    We count the number of Stack Exchange instances per type (with ``E'' standing for \aaemerging (in green), ``T'' for \aatransitioning (in blue) and ``S'' for \aasustainable (in pink)) every six months until the first three years of the instances' existence.
    We highlight that \aasustainable instances take at least a couple of years to develop, and \aaemerging instances typically grow to the \aatransitioning type in less than two years.
    This temporal process suggests \aaacs of Stack Exchange instances shift and mature with time.
}
\end{figure*}
Now, we analyze how a Stack Exchange instance's type (i.e. its \aaac in terms of \aats) changes over time.
To do so, we count the number of instances per type at different points of their existence in Figure~\ref{fig:dataset_type_per_quarter}.
Specifically, starting with the first six months after inception, we categorize each instance over the course of its existence (in increments of six months until three years) as \aaemerging (or ``E'' in Figure~\ref{fig:dataset_type_per_quarter}), \aatransitioning (``T'') or \aasustainable (``S'').
Note that $49$ out of $50$ Stack Exchange instances are at least six months old, but only $32$ are at least three years old.

We note that, after the first six months, almost all instances are of the \aaemerging type.
We observe that there is no Stack Exchange instance which immediately transitions from the \aaemerging to the \aasustainable type.
Most instances need at least 18 to 24 months before moving from the \aaemerging type to the \aatransitioning type.
$15$ out of $32$ (roughly $47$\%) Stack Exchange instances evolve to the \aasustainable type by their third year.

This developmental process suggests the \aaacs we propose correspond to maturity stages of Stack Exchange instances.

\section{Discussion}
We discuss the \aats with respect to their impact, comparable user behavior identified in related literature, and their role in Stack Exchange instance development.
Basing on that discussion, we derive practical implications for Q\&A community managers to optimize their community development efforts.

\noindent \textbf{Impact of \aats in the context of user characterizations in related work.}
We first discuss similarities and key differences in \aats and the two or three types of temporal user activity patterns other authors typically mention in their studies of online Q\&A instances \cite{mamykina2011design, sinha2013exploring, furtado2013contributor}.
The main difference between our \aats of user behavior and others lies in the \aanonrecurring.
When describing least active users, namely low-profile~\cite{mamykina2011design}, low activity~\cite{furtado2013contributor} and less participatory~\cite{sinha2013exploring} ones, these authors group users with both short and long term lurking behavior.
By splitting such lurking behavior into the \aanonrecurring and the \aasporadic, we uncover a distinction in lurking behavior with respect to a fundamental aspect of users' participatory interest in a Q\&A community.
Specifically, users of the \aanonrecurring seem to join a Q\&A community with a specific question or purpose and leave after it is fulfilled.
This behavior poses a contrast to users of other \aats as well as of the low activity types found by other authors:
These other users participate in a Q\&A community for longer periods of time, suggesting a higher interest in the Q\&A community itself or at least in more of its topics.
We argue for our granular characterization of low profile user behavior due to the high impact users of the \aanonrecurring have: (i) They represent the majority of total user base and (ii) their role remains important throughout the development of Q\&A communities' activity dynamics.
We believe other \aats could correspond more directly to user behavior described by these other authors: The \aasporadic could correspond to the occasional~\cite{furtado2013contributor} and partially shooting star~\cite{mamykina2011design} user profiles, the \aafrequent to the answer activist~\cite{furtado2013contributor} and the more participatory users~\cite{sinha2013exploring}, and the \aapermanent to the community activist~\cite{mamykina2011design} and hyperactivist~\cite{furtado2013contributor}.
In agreement with descriptions by these other authors, we see a prominent role by users of the \aasporadic as the least active type of users which at least engages with the community and thereby provides questions and, to a lesser extent, answers, spread out over time. 
A large gap in the activity profiles of the \aasporadic and the \aafrequent makes the difference between them obvious.
We argue for a distinction between \aafrequent and \aapermanent due to the former's high activity profile and non-negligible user count as a backbone of the community, effectively balancing the workload with the relatively few users of the \aapermanent.
Interestingly, Furtado et al.~\cite{furtado2013contributor} describe users with the hyperactivist role as those which even participate in community moderation, supporting our view users of the \aapermanent act as community leaders.
Note, however, that we do not claim any of these correspondences are a perfect match, as each of those authors and ourselves focus on different facets of user activity and behavior in online Q\&A communities.

Regarding our feature choice to describe \aats, our analysis reveals that a simple, small set of features is sufficient for separating temporal user activity patterns.
We see these facts as a promising result for future Q\&A activity dynamics modeling efforts---by using only a small number of parameters, models can be kept simple and interpretable (e.g., we may model user activity as a simple Poisson process), but still effective and accurate. 
Moreover, empirical estimation of parameters for simple models is typically easy and efficient.

\noindent \textbf{Dynamics of \textit{activity compositions}.}
Having reiterated the significance of our \aats characterization, we discuss the importance of their roles at different stages of a Stack Exchange instance's development.

We observed that Stack Exchange instances of the \aatransitioning and \aasustainable type exhibit an oscillating and respectively growing flow of question activity coming mostly from the \textit{Non-Recurring} and \textit{Sporadic Activity Archetypes}.
We thus believe initially setting low entrance barriers and providing incentives for one-time and infrequent external impulses, in the form of participation by the users of the \aanonrecurring and the \aasporadic, as they form the basis for successful activity development of Q\&A communities.
Research~\cite{kittur2007power, suh2009singularity, solomon2014critical} on the roles played by novice users and their activity dynamics in online collaborative communities such as Wikipedia supports our reading regarding the importance of these less active users.
We believe the second ingredient for successful activity development lies in the community's reaction to activity by both less active user archetypes, since both expect answers and comments from the communities they engage with.
We observe, in \aasustainable instances, that users of the \textit{Frequent} and \textit{Permanent Activity Archetypes} bear the bulk of this workload, whereas in \aatransitioning it's the users of the \textit{Non-Recurring} and \textit{Sporadic Activity Archetypes} themselves.
We note that publicly available statistics from Area 51 datasets \cite{tridion2017area51} and previously mentioned work~\cite{mamykina2011design, furtado2013contributor} stress the importance of this core community from the \textit{Frequent} and \textit{Permanent Activity Archetypes}.
Other studies focusing on knowledge sharing dynamics of other online Q\&A communities~\cite{adamic2008knowledge, nam2009questions} even correlate higher activity levels with question answering performance, thus reinforcing the key role the most active users play.
To summarize, our results suggest activity growth-inducing structures prominently feature a core of recurring users, experts and community leaders of the \aafrequent and \aapermanent, and a steady, numerous stream of users of the \textit{Non-Recurring} and \textit{Sporadic Activity Archetypes}.

In \aaemerging Stack Exchange instances, the dynamics of the four main \aats have not formed yet, so other clusters of activity types, not belonging to any of the four archetypes, dominate.
We reason that this is a direct consequence of \aaemerging instances simply lacking users and time required to establish structures to support these user activity dynamics. 

\noindent \textbf{Practical implications for growing Q\&A communities.}
Based on our analysis we propose a series of measures for operators of Q\&A instances to focus on as they grow and foster their communities.

For an operator starting a Q\&A community, our analysis indicates her first priority should be to gather users interested in the community she oversees, possibly via integration with other topic related online communities~\cite{nam2009questions}.
To do so, we suggest the community operator initializes the community in a controlled beta phase, as proposed also by~\cite{mamykina2011design, tridion2017area51}, with the intent of establishing simple sets of rules, which ease the load on operators and moderators and ensures newcomers feel welcome.
Newbie corners and close monitoring of this initial phase, to e.g. continuously improve ease-of-access and not introduce counterproductive overregulation~\cite{suh2009singularity}, should help the community improve activity levels beyond beta status. 
Although Kittur et al.~\cite{kittur2007power} suggest experts were crucial to bring content and utility to the early days of Wikipedia, our results indicate young Q\&A instances, i.e. those less than two years old (cf. Fig.~\ref{fig:dataset_type_per_quarter}), also benefit strongly from bursty activity by users of the \aanonrecurring and \aasporadic.

This does not imply, however, that the \aafrequent and \aapermanent should be neglected, as developing and rewarding recurrent participation in a Q\&A becomes more important over the mid-term of $18$ to $36$ months (again, cf. Fig.~\ref{fig:dataset_type_per_quarter}).
In this phase, community operators could invest in a badge and gamification system to elicit more participation and community spirit by users of the \aafrequent and \aapermanent, as these badges and gamification elements have been shown to enhance participation by these types of users~\cite{nam2009questions, anderson2013steering, immorlica2015social, easley2016incentives, zhang2016social}.
Furthermore, community question routing systems, such as the one proposed by Srba et al.~\cite{srba2015utilizing}, should help matching questions to answerers, thus ensuring the needs of the users of the \aanonrecurring and \aasporadic are met.
Finally, operators should gather feedback from their user base continuously~\cite{mamykina2011design} and engage leaders, potentially such as those of the \aapermanent, to help with community moderation~\cite{furtado2013contributor}.

\section{Conclusions}

In this paper, we uncover temporal activity patterns in 50 Stack Exchange Q\&A instances at both the user and instance levels.
To achieve this, we start by representing user activity in those instances as time series, which comprise the total count of users' questions and answers over time.
We extract representative features from these time series to better cluster them and to derive an optimal numbers of clusters.
These clusters represent a set of four \aats, which characterize users mainly according to the frequency of participation in a Q\&A community. 
Then, we break down activity in Stack Exchange instances by the different \aats, which allows us to recognize three instance types: \aasustainable, \aatransitioning and \aaemerging.
\aasustainable instances have the highest levels of activity and the largest number of active users.
Their success correlates with a small but strong backbone of users of the \textit{Frequent} and \textit{Permanent Activity Archetypes}, reacting to a steady flow of users from the \textit{Non-Recurring} and \textit{Sporadic Activity Archetypes}.
We find that \aaemerging and \aatransitioning Stack Exchange instances either completely lack or are in the process of establishing such activity profiles.
Our \aats and Stack Exchange instance characterization allow us to measure online Q\&A instance health and success.
We provide a methodology for community managers of Q\&A instances to detect the maturity stage of their communities, and we recommend activity composition structures for them to aim for, as well as concrete steps to take to help their communities mature from one stage to the next.

Besides the aforementioned limitation regarding feature selection and corresponding clustering quality and interpretation (as other binary features might yield equally good clustering quality but other interpretations), we reflect on the generalization and practical implications of our approach with respect to other Q\&A datasets.
Although our proposed features are suitable for capturing general bursty types of activity found in Q\&A instances, these features might need tailoring in the application to Q\&A communities besides Stack Exchange instances.
In particular, the threshold for the feature based on the number of activity peaks will vary depending on the Q\&A platform, which is why we defined it as a data-dependent percentile value.
Moreover, the choice of granularity of time series aggregation, in our case monthly, must be taken with care, since too coarse a temporal resolution will hide burstiness and activity peaks, and too granular a resolution will lead to time series with longer periods of inactivity and thus a less distinguishable ``ratio of unique non-zero values'' feature.
However, once time series granularity and our proposed features have been adjusted for a potentially new dataset, we expect the clustering to yield comparable results, since our proposed features yield clear-cut separated clusters.
Therefore, we expect practitioners working with our proposed approach to be able to gauge their extension to their datasets, in particular in case of modifications to our proposed features, by evaluating the resulting clustering quality and checking if it is comparable to the one we report.
One last noteworthy limitation regards the fact the Stack Exchange instances we analyzed do not become completely inactive. As such, we refrain from discussing the generalization of our proposed approach in the case of ``death'' of Stack Exchange instances.

Naturally, empirically verifying the generalization of our method to other Q\&A platforms would be of great interest. Moreover, conducting small-scale real experiments would further cement our argumentation on this work's practical implications.
Other future work includes mathematical modeling of activity in online Q\&A communities based on the \aats and their activity compositions with the aim of deriving further recommendations for operators to assess and optimize their online presence.
Finally, enhancing our analysis to include quality-related aspects of activity in Q\&A communities would be of great interest.

% Bibliography
\bibliographystyle{ACM-Reference-Format}
\bibliography{sample-bibliography}

\end{document}